\documentclass[fleqn,usenatbib]{mnras}

\usepackage{newtxtext,newtxmath}
\usepackage[T1]{fontenc}
\usepackage{ae,aecompl}
\usepackage{hyperref}
\usepackage{graphicx}	
\usepackage{xcolor}

\usepackage{amsmath}	
\usepackage{scalefnt}
\usepackage{threeparttable, tablefootnote}
\usepackage{wrapfig}	
\usepackage[textwidth=1.5cm,shadow]{todonotes}	
\usepackage{float}
\usepackage{color, soul}	
\usepackage{verbatim}	
\usepackage{fontawesome}    
\usepackage[normalem]{ulem} 

\usepackage{lineno}

\newcommand{\orcid}[1]{\href{#1}{\includegraphics[scale=0.035]{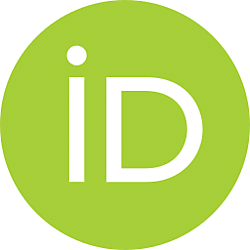}}}

\definecolor{boxback}{HTML}{DAE5F0}
\definecolor{boxframe}{HTML}{0F365B}
{
\end{tcolorbox}   
}

\newcommand{\eg}[1]{(e.g. \citealt{#1})}

\def\code#1{\texttt{#1}}

\def\quote#1{``#1''}

\title[Black Hole Deep Learning]{Black Hole Weather Forecasting with Deep Learning: A Pilot Study}

\author[Duarte, Nemmen \& Navarro]{
Roberta Duarte$^{1}$\thanks{E-mail: roberta.duarte@usp.br},
Rodrigo Nemmen$^{1}$\thanks{E-mail: rodrigo.nemmen@iag.usp.br}\orcid{https://orcid.org/0000-0003-3956-0331}, 
João Paulo Navarro$^{2}$ 
\\
$^{1}$Universidade de S\~ao Paulo, Instituto de Astronomia, Geof\'{\i}sica e Ci\^encias Atmosf\'ericas, Departamento de Astronomia,\\ S\~ao Paulo, SP 05508-090, Brazil \\
$^{2}$NVIDIA
}

\date{Accepted 2022 March 3. Received 2022 February 10; in original form 2021 January 11}

\pubyear{2022}

\begin{document}
\label{firstpage}
\pagerange{\pageref{firstpage}--\pageref{lastpage}}
\maketitle

\begin{abstract}
In this pilot study, we investigate the use of a deep learning (DL) model to temporally evolve the dynamics of gas accreting onto a black hole in the form of a radiatively inefficient accretion flow (RIAF). We have trained a convolutional neural network (CNN) on a dataset which consists of numerical solutions of the hydrodynamical equations, for a range of initial conditions. We find that deep neural networks trained on one simulation seem to learn reasonably well the spatiotemporal distribution of densities and mass continuity of a black hole accretion flow over a duration of $8\times 10^4 GM/c^3$, comparable to the viscous timescale at $r=400 GM/c^2$; after that duration, the model drifts from the ground truth suffering from excessive artificial mass injection. Models trained on simulations with different initial conditions show some promise of generalizing to configurations not present in the training set, but also suffer from mass continuity issues. We discuss the caveats behind this method and the potential benefits that DL models offer. For instance, once trained the model evolves a RIAF on a single GPU four orders of magnitude faster than usual fluid dynamics integrators running in parallel on 200 CPU cores. We speculate that a data-driven machine learning approach should be very promising for accelerating simulations of accreting black holes.
\end{abstract}


\section{Introduction}

Black holes (BHs) are infinitely deep gravitational potential wells surrounded by event horizons--surfaces that separate the outside world from the region of the BH from which nothing escapes. When matter falls into such a hole it forms a disk-like structure due to the barrier posed by angular momentum conservation; given this barrier, magnetic forces in an ionized plasma supply the friction required to allow gas to fall onto the BH \citep{Balbus2003}. These stresses also convert some of the gravitational potential energy of the accretion flow into heat and can release a substantial fraction of its rest mass, providing the primary power source behind active galactic nuclei (AGNs), black hole binaries and gamma-ray bursts \citep{Meier2002}. 

At the same time that magnetic stresses make BHs shine through the release of electromagnetic radiation, they also generate turbulence in the accretion flow thereby turning accreting BHs into test beds of fluid dynamics. The nonlinear partial differential equations that need to be solved in order to describe the  turbulence, gravity and radiation in the spacetimes around BHs are intractable analytically. The traditional approach to deal with such a problem has been to numerically solve the partial differential equations behind the conservation laws for the system \eg{Gammie2003, Mignone2007, Toro2009}. With the numerical solutions, one can then perform detailed studies of the multidimensional gas dynamics and radiative properties of BH accretion flows---i.e. BH weather forecasting---being only limited by the available computational resources. Such numerical simulations have been a key aspect in  providing a framework for interpreting the multitude of observations of black holes and their environments \eg{EHTC2019d}. 

Currently, a hydrodynamical 3D model with a competitive spatial resolution (e.g. $N_r \times N_\theta \times N_\phi = 400 \times 200 \times 200$ cells) requires about $6 \times 10^7$ CPU-hours to be evolved for a duration of $10^5 GM/c^3$ (e.g., \citealt{Almeida2020}; hereafter, we adopt units such that $G=c=1$, i.e. both $G/c^3$ and $G/c^2$ are unity). Therefore, a scientist that desires to reproduce or improve upon such a model needs to have access to a CPU cluster with thousands of cores. The computational cost can be reduced by a factor of $\approx 15$ if one adopts a code optimized for graphical processing units (GPU; \citealt{Grete2019} for the Newtonian magnetohydrodynamical case, \citealt{Liska2019b} for general relativistic magnetized one), but one still would need to have access to a GPU cluster. Large computational costs are of course not an exclusive issue of BH astrophysics---they are also a problem in many other fields. One particular example are cosmological simulations of large scale structure formation where one needs to evolve the gravitational assembly of dark matter haloes and their baryonic physics \eg{Vogelsberger2014, Schaye2015}. 

A recent innovation is using machine learning (ML) as an approach to computational simulations of large spatiotemporally chaotic systems \citep{Brunton2020}. The basic idea is to use the tools of ML to ``learn from experience'': instead of directly simulating the physical processes involved, devise the prediction as a computer vision problem and infer the evolution of the system from the sequence of input data cubes that comprise previous states. In other words, this is a data-driven, physics-free approach in which a ML model learns to approximate the physics from the training examples alone and not by incorporating a priori knowledge about the equations underlying the processes \eg{Jaeger2004}.

Deep learning (DL) is a particular class of ML models which is proving quite promising for data-driven forecasting of complex systems. DL is a type of supervised learning, where the model is trained (or fitted in astronomical jargon) with many input-output examples \citep{LeCun2015}. For example, given many galaxy images labeled as elliptical or spiral, learn to predict whether a given image is that of a elliptical or spiral \eg{Hausen2019}. This is achieved by updating the weights of a multi-layered (i.e. deep) neural network (NN) via gradient descent with a differentiable loss function. The flexibility of deep nets renders DL a good approximator for functions which are too complex to have an analytical form \citep{Cybenko1989,Hornik1991,Zhou2020}. For DL to produce acceptable results, a large amount of training data is essential \eg{LeCun2015,Krizhevsky2017}.


We review a couple of exciting applications of this approach for data-driven forecasting of nonlinear systems. 
\cite{Tompson2016} used a convolutional neural network (hereafter CNN) combined with an unsupervised learning framework to learn the 3D solutions of the inviscid Euler equation. Their data-driven simulations outperform other methods and show good generalization properties. Similarly, \cite{King2018,Mohan2019} obtained promising results using a long term short term CNN architecture. \cite{Pathak2018} employed a reservoir computing paradigm to forecast the solutions of the chaotic, Kuramoto-Sivashinsky equation over the large duration of six Lyapunov times. Using deep nets without convolutions, \cite{Breen2019} were able to forecast with good accuracy the 3-body problem; \cite{Battaglia2016} did something similar with CNN for N-body systems with $N<20$. \cite{Agrawal2019, Chattopadhyay2020, Ravuri2021} used different DL methods for traditional Earth-weather forecasting, with promising results. Recently, \cite{kochkov2021} proposed an end-to-end deep learning to accelerate computational fluid dynamics by improving approximations.

It is noteworthy that there are works using deep learning applications to solve problems around black hole physics. \cite{Lin2020} proposed using CNNs to obtain black hole parameters from synthetic images obtained from simulations. They fed synthetic images to a traditional CNN with a multi-layer perceptron (MLP) attached to output the spin and magnetic flux from the system. Their model resulted in a high accuracy method that can recover spin and magnetic flux.  Similarly, \cite{Gucht2020} proposed two neural networks, one to obtain parameters such as viewing angle, position angle, mass accretion rate, electron heating, and the black hole mass, the second to obtain the spin. They fed synthetic images from black hole shadows and obtained a model that accurately recovered mass and accretion rate. Finally, \cite{Lin2021} also proposed a data-driven model that can recover parameters from interferometric data and obtained a method to recover magnetic flux without reconstructed images.

In this work we address two interrelated questions. The first question is a fundamental one: can DL learn and forecast the hydrodynamical evolution of astrophysical systems? Concretely, how good is it in predicting the future of a spatiotemporally chaotic system comprised by a turbulent fluid? If it is able to forecast the future, for how long is the quality of the forecast acceptable? The second question is one of practical order: is it possible to obtain acceptable solutions of the fluid conservation equations (such as the Navier-Stokes equation) faster than an explicit numerical solver using DL techniques? 

This work is a pilot study of DL techniques applied for BH weather forecasting. The astrophysical setting which provides the training dataset for our study consists of a hydrodynamical simulation of a BH accreting gas from a geometrically thick torus of very hot gas, also called a radiatively inefficient accretion flow (RIAF; e.g. \citealt{Yuan2014, Almeida2020}). RIAFs are thought to be the most common mode of BH accretion in present-day galaxies \eg{Ho2008, Nemmen2014, EHTC2019}, where BHs are accreting at mass accretion rates $\dot{M} < 0.01 \dot{M}_{\rm Edd}$ where $\dot{M}_{\rm Edd}$ is the Eddington accretion rate. For this reason, there is wide interest in modelling RIAFs by numerically solving the conservation laws \eg{Porth2019}.

For the learning algorithm, we use the well-known U-Net architecture \citep{Ronneberger2015}, which is commonly used to extract patterns from datasets with spatial (images) and temporal (videos) coherence \eg{Karpathy2014, Chen2016}.
One example of a successful application using the U-Net architecture is the TF-Net \citep{ruiwang2020} for turbulent flows in the absence of gravity. \cite{ruiwang2020} compared the performance of TF-Net with different architectures and they found that U-nets have superior performances. \cite{chen2020} and \cite{kim2019} showed that generative models can also be suited to predicting the evolution of such flows using a data-driven approach. \cite{Pfaff2020} presented an experiment by predicting the states of several physical systems using graph neural networks, including predicting the states of a fluid. Another successful case is using hybrid components, i.e., combinations of different approaches, such as \cite{cheng2020} where a hybrid model using generative models with variational autoencoders is applied to model fluid flows.

Our motivation for choosing the U-Net architecture is the following. Firstly, U-net preserves the spatial relations present in the data set \citep{Ronneberger2015}. Second, the U-Net preserves critical information while the data pass through the architecture since it has skip connections between the encoder and decoder. The information (e.g., spatial and temporal) that may be lost in the encoder process can be recovered through skip connections. The skip connections force the decoder to consider the input and encoder outputs while making a prediction. A drawback of using U-Net is that it does not save temporal information as in recurrent neural networks \citep{giles1994}. We overcame the issue by modifying the architecture to receive and treat temporal information in the fourth dimension of the tensor.

The BH simulation generates the spatiotemporal distribution of the density field which we feed to the DL model. The challenge then is how well the DL model predicts the future state of the density field. We quantify the performance of the DL physics-free approach by comparing a number of indicators with those obtained from the explicit solution to the conservation equations. 

The paper is organized as follows: In \hyperref[sec:data]{
section 2} we present the fiducial equation-based model that we used to generate the training data and the features. In \hyperref[sec:methods]{section 3} we describe the DL model and the custom loss function that we developed. In \hyperref[sec:results]{section 4} we describe the results. In section \ref{sec:disc} we present the discussion and finally, in section \ref{sec:conclusion} we finish with the conclusions of the work.

An aside about the nomenclature. From here onwards, when we refer to a \quote{model} we are referring to the DL model trained with the data from the hydrodynamical simulation. A \quote{frame} corresponds to an instance of the training dataset (as in a movie frame).

\section{Data}  \label{sec:data}  


Our data-driven approach is supervised, meaning that for each training example $x$ we  provide the correct answer $y$. The ML model then learns the relation $x \rightarrow y$ that best fits the provided training examples. Our goal is to first provide a meaningful set of examples to teach the model about the physics represented in the data. 

In the ML context, the term ``learn'' has a straightforward meaning: train a model on some data and deliver predictions $\hat{y}$  (the learned model's output) as close as possible to the ground-truth $y$ (the data). If the trained model is successful, it would generalize well---i.e. strong generalization in this context would imply reproducing the spatiotemporal evolution of a BH accretion flow simulation even for  initial conditions that are not present in the training dataset.

\begin{figure}
\centering
\includegraphics[scale=0.45]{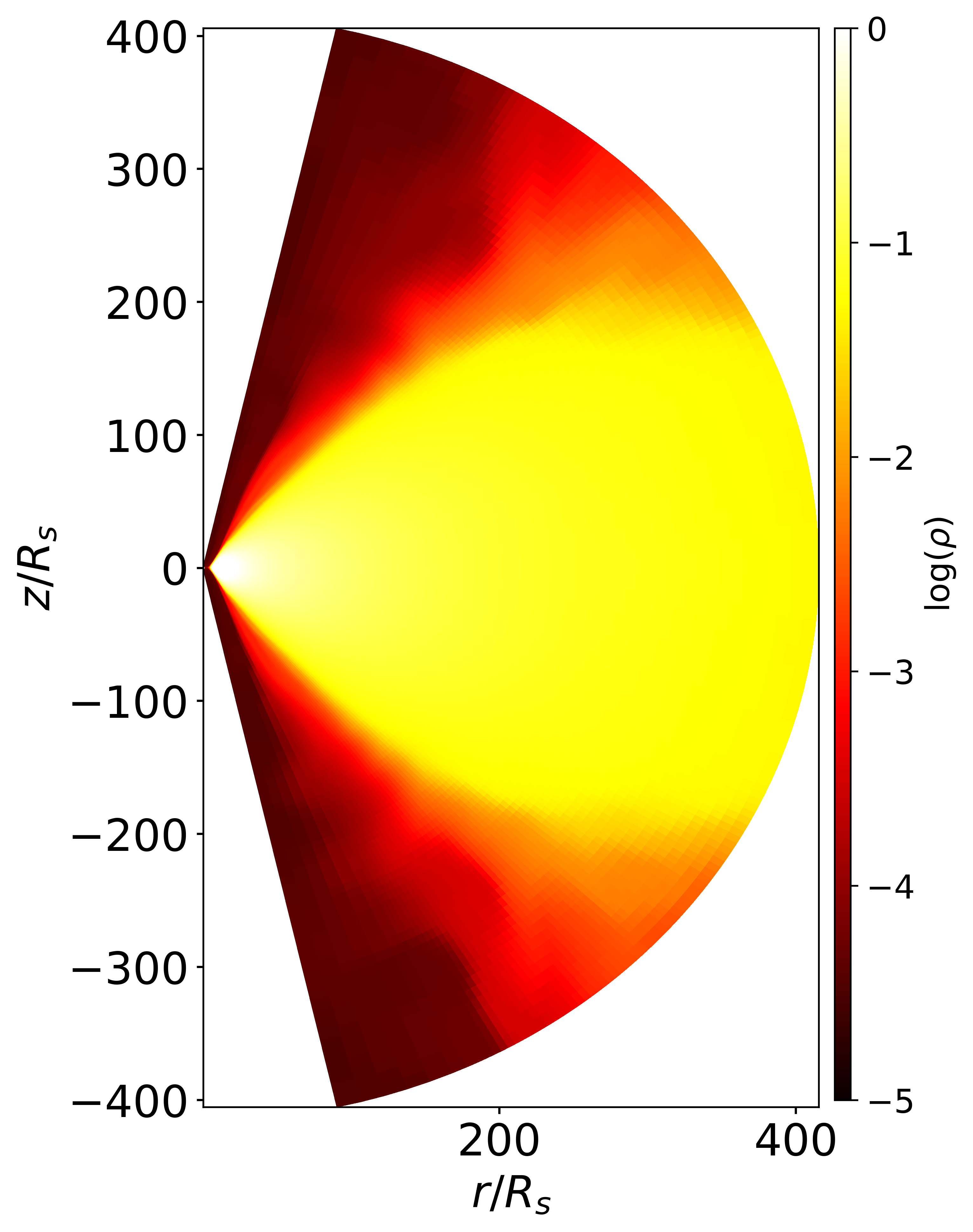}
\caption{Simulated gas density (logarithm) around a black hole at $t = 257387 M$, which comprises one example of the training set. The lengths are in units of the Schwarzschild radius, $R_s \equiv 2GM/c^2$.}
\label{fig:sim-example}
\end{figure}

Concretely, we are interested in training a model to reproduce the density field $\rho(\textbf{r}, t)$, which is the feature used to find the best model (cf. Figure \ref{fig:sim-example}). Our dataset was generated from two-dimensional hydrodynamical simulations of viscous accretion onto a Schwarzschild BH (\citealt{Almeida2020}, hereafter AN). The BH gravity was approximated with a pseudo-Newtonian potential which reproduces many of the Schwarzschild geometry properties \eg{Abramowicz2009}. The simulations were performed using the code \code{PLUTO} which employs a Godunov-type scheme to solve the fluid  equations \citep{Mignone2007}, namely:
\begin{align}   \label{navierstokes}
&    \frac{d\rho}{dt} + \rho \nabla \cdot \textbf{v} = 0, \\
&   \rho \frac{d\textbf{v}}{dt} = \nabla P - \rho \nabla \psi + \nabla \cdot \textbf{T}, \\
&    \rho \frac{de/\rho}{dt} = -P \nabla \cdot \textbf{v} + \frac{\textbf{T}^2}{\mu}, 
\end{align}
where $\rho$, $\textbf{v}$, $P$, and $e$ are the density, velocity, pressure, and internal energy, respectively. The pseudo-Newtonian potential is given by $\psi = GM/(r - R_s)$. Angular momentum is removed from the accreted gas via magnetic stresses, which are modelled using an effective prescription called ``$\alpha$-viscosity'' (cf. AN for more details).  

The initial condition corresponds to a rotating torus in dynamical equilibrium extending from an initial radius of $10\;-\;40 M$ to a final radius of $1000M$. AN explored two different prescriptions for the angular momentum distribution $l(R)$: (i) A power-law distribution $l(R) \propto R^a$ with $a$ varying in the range $0 - 0.4$ which we will refer to as ``PL'' setup; (ii) The distribution proposed by \citet{Penna2013}, namely 
\begin{equation}
l(R) = \begin{cases}
    {\rm constant}       & \quad R < 21 R_S \\
    0.71l_K  & \quad \text{otherwise} 
\end{cases} 
\end{equation}
where $l_K$ is the Keplerian specific angular momentum. We refer to this prescription as ``PN''. Similarly, the viscosity can also be separated into two prescriptions: (i) a prescription called "K-model" in \cite{Stone1999}, $\nu = \alpha r^{1/2}$, we will refer this as ST. (ii) a parametrization from \cite{Shakura1973}, $\nu = \alpha c_s^{2}/\Omega_K$, and it will be referred as SS. In both cases, the parameter $\alpha$ can have values of 0.1, 0.3, and 0.01. 

The calculations are performed in a fixed mesh extending up to $10^4 R_s$ with a resolution of $400 \times 200$ cells in polar coordinates ($N_r \times N_\theta$). We use a non-uniform mesh with higher resolution towards smaller radii $r$. Regarding the computational mesh as an image is a novelty from the point of view of computer vision DL applications, since they usually represent images or videos using regular, uniform grids. 

The inner and outer boundary conditions correspond to ``outflow'' boundaries, where all gradients are zero. A total of nine simulations were performed with durations ranging from $8 \times 10^4 M$ to $8 \times 10^5 M$ which are extremely long for today's standards and comparable to the viscous time at the midpoint of the torus. The simulations differ in the their angular momentum distribution and amount of shear stress. In this work, the density maps from the simulations are the training dataset for the ML algorithms.


\section{Machine learning methods} \label{sec:methods}

\subsection{Convolutional neural network}

Our NN architecture is based on the U-Net, proposed by \citet{Ronneberger2015} (cf. Appendix \ref{sec:arch} for details on the architecture). We feed the model with a $3$--dimensional array composed of the $R \times \theta \times t$ density coordinates. The network is composed of an encoder and a decoder. While the encoder maps the input into the latent space where the data are mapped into a compressed representation, the decoder maps the latent space into another array which is the output (i.e. the forward pass) of the DL model. 


Since our data present spatial correlations, we adopt convolutional layers which capture these correlations. Convolutional layers put together in a sequence result in a convolutional neural network (CNN). CNNs are powerful tools to solve problems in which the data present spatial and temporal coherence, where classical multi-layer perceptron \citep{Murtagh1991} approaches fail either for the absence of numerical accuracy or due to the high-computational complexity. 


Here we give a summary of how a CNN works. For in-depth accounts please refer to \cite{LeCun2015, Goodfellow2016}. The free parameters of a CNN are  elements of a convolution filter. In the forward pass, we perform successive convolutions  followed by non-linear activation functions until we reach the last layer, which produces the prediction $\hat{y}$. An error function $\mathcal{L}$---which is called loss function in the ML community---measures the difference between $\hat{y}$ (the prediction from the DL model) and the target value $y$ (the data or ground-truth). As usual in statistical model fitting, we want to find the values of the parameters that minimizes $\mathcal{L}$. Here, the parameters are the weights in the convolution filters. 

The process of finding $min(\mathcal{L}(x; w))$ ($x$ represents the data) is a non-convex optimization problem, with no warranty of global minima. We adjust the weights iteratively, in a stochastic manner until we find a set of parameters that best minimizes the loss function for a given training dataset. We have the freedom of choosing the loss function as any statistic that conveys the difference between $\hat{y}$ and $y$. Common choices for $min(\mathcal{L})$ are the mean absolute error and mean absolute percentage error.

ML algorithms have other free parameters called hyperparameters which are used to control the learning cycle, i.e. the number of iterations, convergence criteria and network architecture. The hyperparameters vary freely and are independent of each other. They strongly impact the the quality of the trained model. We used a grid search method combined with a random search method \citep{bergstra2012} to find the best combinations. Our method evaluates several combinations of hyperparameters with random values, using the coefficient of determination $R^2$ as an error metric in the validation set, whose minimization gives the best values of the hyperparameters. 

We performed the data preparation steps detailed in Appendix \ref{sec:dataprep} before we feed the algorithm with the frames. The input is a block $(1, 256, 192, 5)$, and the output is the block equivalent to five frames ahead. The channels in the blocks represent five consecutive frames. Our CNN is trained with blocks composed of $(64, 256, 192, 5)$, where $64$ is the batch size.

\subsection{Loss function}

A suitable loss function ensures a good convergence of the learning procedure. One contribution of our work is the definition of $\mathcal{L}$ adapted to the challenge of dealing with a density that varies spatiotemporally as is natural for BH accretion, where the density increases towards smaller radii. 

\begin{figure}
\begin{flushleft}
\includegraphics[scale=0.2]{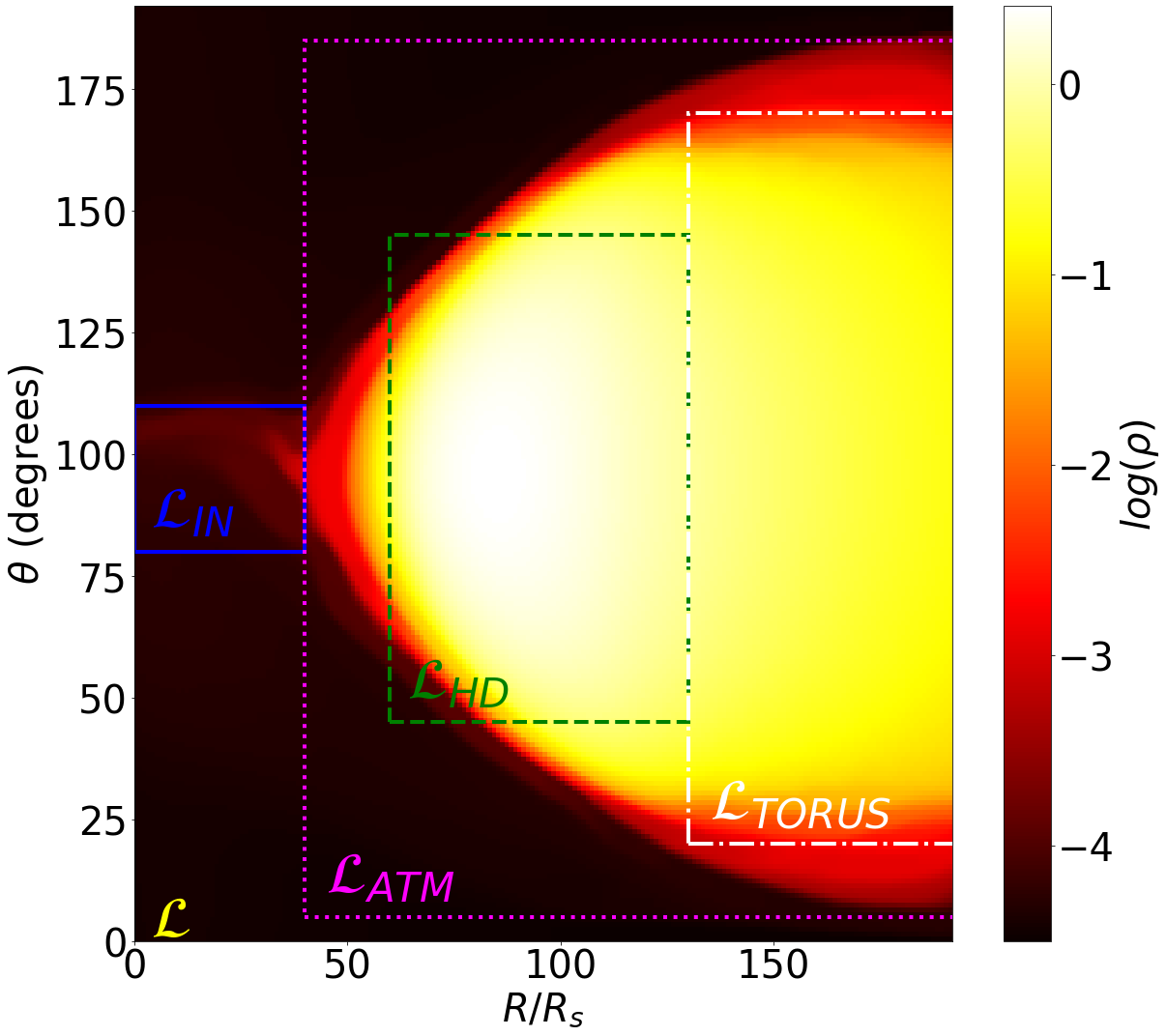}
\caption{Schematic diagram illustrating the different regions of the flow where the components of the hierarchical loss function are calculated. In each box, we calculate the difference between the prediction $y$ and the target $\hat{y}$ of the marked region. The total loss $\mathcal{L}$ is the sum of all these contributions. }
\label{fig:loss}
\end{flushleft}
\end{figure}

When the NN is trained with many examples sharing the same features, it may suffer from bias towards the features that are over-represented in the training set. In our case, this takes the form of the over-representation of density regions with $\rho < 10^{-4}$ in the flow (in code units) 
which occur in a larger volume  than regions with $\rho > 10^{-4}$. To account for this bias, we propose a hierarchical loss function built upon separate functions whose weights vary depending on the region of the accretion flow encompassed, written as 
\begin{equation}
\label{loss}
   \mathcal{L} = \mathcal{L}_{\text{T}} + \alpha\mathcal{L}_{\text{HD}} + \beta\mathcal{L}_{\text{IN}} + \gamma\mathcal{L}_{\text{TORUS}} + \delta\mathcal{L}_{\text{ATM}}.
\end{equation}
The convention for each loss function component building the total value $\mathcal{L}$ is the following: $\mathcal{L}_{\text{T}}$ is the loss for the entire flow, $\mathcal{L}_{\text{HD}}$ for the high-density region (the region with $\bar{\rho} > 0.9$ where $\bar{\rho}$ is the mean normalized density), $\mathcal{L}_{\text{IN}}$ for the inner regions (the region with $\bar{\rho} < 0.1$), $\mathcal{L}_{\text{TORUS}}$ for the torus (the region with $\bar{\rho} = 0.5$) and $\mathcal{L}_{\text{ATM}}$ for the atmosphere (the region excluding where ${\rho} < 0.05 $). Figure \ref{fig:loss} shows the regions chosen where the functions just described are calculated. The weights $\alpha$, $\beta$, $\gamma$ and $\delta$ correspond to hyperparameters. We choose the losses as mean absolute errors since the absolute error is robust when dealing with outliers, except for $\mathcal{L}_{\text{T}}$ which is a mean squared error. We considered other losses for our problem, however the choice reflected in equation \ref{loss} returned the best results. 

Our loss was defined by analyzing the behavior of a metric --- in this case, $R^2$ as we will discuss in section 3.4 --- evaluated in the validation set, after the training procedure. We performed different trainings considering several combinations of losses and focused on the training resulting in the best metric value (for $R^2$, 1 is the best value). We visualized how the model was learning each region by calculating the mean squared error between target and prediction of the validation set. By quantifying how much the model learned each region, we could set a weight to the loss representing each torus' region.

The relative size of the regions used to define the components of the loss in equation \ref{loss} is fixed in time and chosen by eye. This is appropriate when performing training on the dataset comprising a single simulation (hereafter called ``one-sim'' case). When the dataset comes from multiple simulations with different initial conditions (hereafter named ``multi-sims'' case), we use a different loss given by
\begin{equation}
\label{secondloss}
   \mathcal{L} = \mathcal{L}_{\text{T}} + a\mathcal{L}_{\text{H}} + b\mathcal{L}_{\text{L}},
\end{equation}
where $a$, $b$ are hyperparameters and $\mathcal{L}_{\text{H}}$ and $\mathcal{L}_{\text{L}}$ are the losses computed for all regions with $\rho > 0.8$ and $\rho < 0.8$, respectively. Table \ref{table:Table1} presents the results of the best hyperparameters found. 

\begin{table}
\begin{center}
\begin{tabular}{|c|c|}
\hline
Hyperparameter & Value \\ \hline
Batch Size & 64 \\ 
Learning Rate & $5 \times 10^{-4}$ \\ 
$\alpha$ &  8  \\ 
$\beta$ &  5  \\ 
$\gamma$ & 10  \\ 
$\delta$ & 4  \\ \hline
\end{tabular}
\caption{The best hyperparameters are found using the grid and random search methods.}
\end{center}
\label{table:Table1}
\end{table}

\subsection{Numerical experiments}  \label{experiments}

We performed two numerical experiments. In the \textit{one-sim} experiment, we train the CNN on one of the longest duration numerical simulations, \code{PNSS3}, in order to quantify the ability of the network to learn from a single simulation. The model was trained using $70\%$ of \code{PNSS3}'s data, of which $10\%$ is used as validation to obtain the hyperparameters. The remaining $20\%$ is used as the ground truth $y$ which is compared with the model's prediction $\hat{y}$.

We split the data as a function of time. $70\%$ of the training set consists of frames ranging from $257361 GM/c^3$ to $628555 GM/c^3$; the validation set ranges from $628752 GM/c^3$ to $681809 GM/c^3$ and the test set goes from $682006 GM/c^3$ to $787920 GM/c^3$. This division prevents the model from overfitting since the test set is composed of future frames, i.e. data that the trained model is intended to predict. However, we acknowledge that the "one-sim" case looks laminar with slight changes between the frames. We overcame this issue in the "multi-sim" case including data displaying more variability and initial tori with considerable distinctions between them.

In the \textit{multi-sim} experiment, we train the DL model using data from eight simulations in order to evaluate the generalization power of the network. We exclude \code{PL0SS3} from the training. The data preparation for each simulation is the same as for \code{PNSS3} in the one-sim case. We test the \textit{multi-sim} predictions against the dataset \code{PNST1} which displays more variability compared to most of the simulations. We match the number of snapshots in each simulation in order to avoid bias towards any model with longer duration.

\subsection{Evaluation} 

We train our models using $70\%$ of the data as training set, and $10\%$ and $20\%$ for the cross-validation and test sets, respectively.
We have 2678 and 5015 frames to train the one-sim and multi-sim models, respectively. In the multi-sim case, we assess the performance of the learned model against the \code{PL0SS3} simulation, which has 709 frames and is not used in the training. We use the first 250 frames to quantify the generalization power of the model. We also test the multi-sim model against parts of the PNST1 data which was not used in the training.

\begin{table}
\begin{center}
\begin{tabular}{|c|c|c|c|}
\hline
Case & Set & Number of Frames & Time ($GM/c^3$)\\ \hline 
One-sim & Total & $2678$ & $530164$\\ 
One-sim & Training & $1875$ & $371194$\\ 
One-sim & Cross-validation &  $268$ & 53056  \\ 
One-sim & Test &  $535$ & $105914$ \\ 
\hline 
Multi-sim & Total & $5015$ & $992820$ \\ 
Multi-sim & Training & $3511$ & $695073$ \\ 
Multi-sim & Cross-validation &  $351$ & $69487$ \\ 
Multi-sim & Test &  $1153$ & $228259$  \\ 
\end{tabular}
\caption{The frame's intervals of each set from the one-sim case and multi-sim case. In the first column, we indicate the case -- one-sim or multi-sim --. In the second column, we indicate the set's name. In the third and fourth columns, we show the number of frames and the time duration.}
\end{center}
\label{table:frames}
\end{table}

\begin{figure}
\centering
\includegraphics[width=\linewidth]{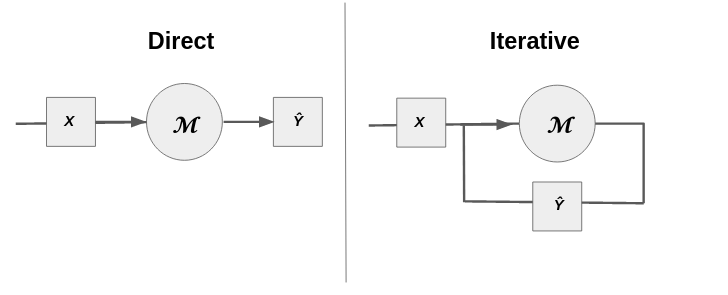}
\caption{The two types of forecasts used in this work, direct (left panel) and iterative (right panel). The $\mathcal{M}$ is the model already trained, $x$ is the snapshot from simulation, and $\hat{y}$ is the prediction. In the iterative forecast, we feed $x$ as the first step and then compute the predictions by iteration.}
\label{fig:forecasts}
\end{figure}

There are two types of forecasts that we perform using the DL model. The \textit{direct forecast} consists of the DL model computing a prediction for the immediate next step once fed with a single input simulation frame from the hydrodynamical simulations. The \textit{iterative forecast} consists of iterative computations of the DL model on top of its own output. In other words, in both approaches the learned model receives the input state only once; the direct forecast evaluates how well the learned model advances in time for one time step, similar to the short-term forecasting in meteorology known as nowcasting. The iterative forecast is representative of longer-duration forecasting. Figure \ref{fig:forecasts} illustrates both types of forecasting.

To evaluate the quality of the DL model forecasts, it is useful to quantify the difference between the target data provided by the hydrodynamical simulation (i.e. the ground truth) and the DL model forecast. For this purpose, we compute the difference between densities in the logarithmic space as $\Delta \log \rho \equiv \log T - \log P$ where $T$ and $P$ are the target and learned model prediction density arrays. Spatial averages are denoted by the usual bar above the corresponding variable. These averages are performed adopting a flat spacetime since our training data was generated from Newtonian simulations. In addition to $\Delta \log \rho$ as a measure of the quality of the DL model forecasts, we also use the root mean squared error (RMSE) typical of ML studies:
\begin{equation}
{\rm RMSE} = \sqrt{\frac{1}{N \times M} \sum_i (P_{ij} - T_{ij})^2 }
\end{equation}

\noindent Both metrics were used to quantify the model's performance by analyzing the predictions' quality compared to the target. The best result would be $\Delta \log \rho = 0$ and RMSE $= 0$ since this imply $T = P$. Our goal is to find values of $\Delta \log \rho $ and RMSE converging to $0$. However, it is convenient to analyze the performance during the training using $R^2$, where  $R^2$ is defined as:

\begin{equation}
\label{r21d}
    R^2 = 1 - \frac{\sum_{ij} \left( P_{ij} - T_{ij} \right)^2}{\sum_{ij} \left( P_{ij} - \overline{P_{ij}}\right)^2}.
\end{equation}

\section{Results}   \label{sec:results}

\subsection{Model trained on one simulation}  

\begin{figure*}
\includegraphics[width=\linewidth]{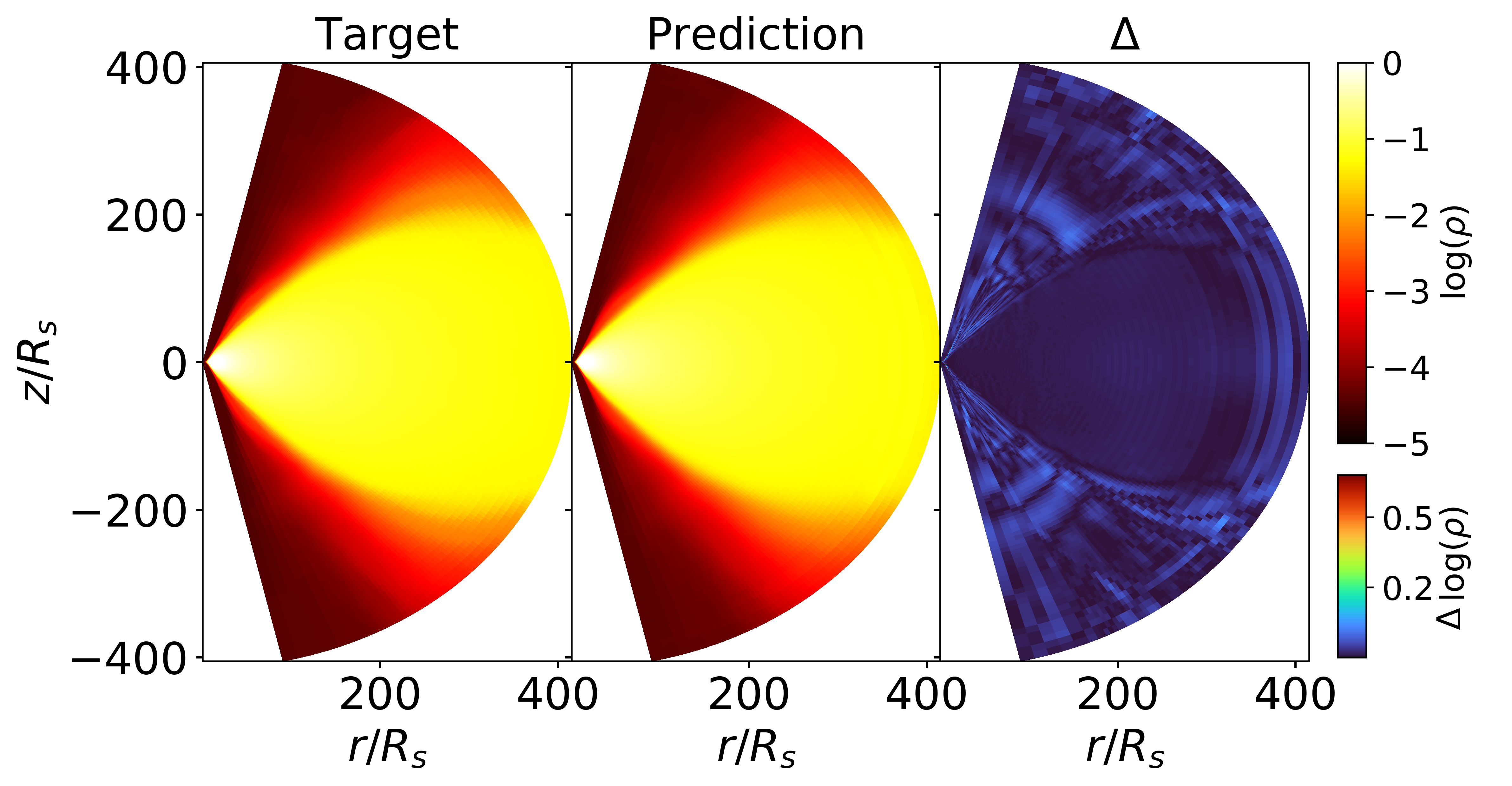}
\caption{Density comparison between the training dataset (left panel) and the DL model \textit{direct} prediction (middle panel) at $t = 607570 M$ for the \textit{one-sim} case. The color maps the logarithm of density. The right panel shows the difference between target and prediction.  } 
\label{fig:direct1}
\end{figure*}

We use the last test set snapshot of PNSS3 at $t = 607570M$ as an input to the trained model. Figure \ref{fig:direct1} compares the prediction from the DL model with target data for the frame that immediately follows, using the direct method in the one-sim case. The density predictions inside the torus (i.e. larger density region) are up to ten per cent different compared to the target data (${\rm max}(\Delta \log \rho) < 0.1$). The predictions are less accurate in the lower density regions (the atmosphere) outside the torus. For instance, the discrepancy between the target and prediction in the atmosphere reaches up to $\Delta \log \rho = 0.2$. 

\begin{figure}
\includegraphics[width=\linewidth]{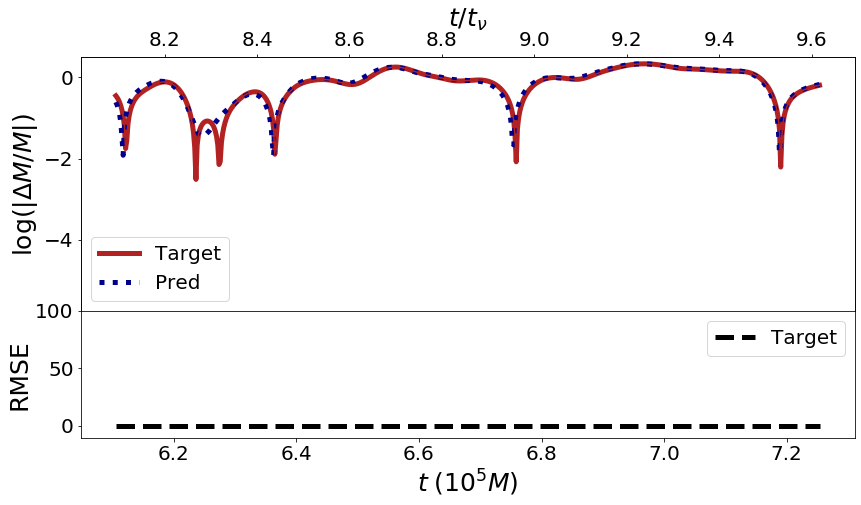}
\caption{Evolution of mass fluctuations based on direct predictions, one-sim model. Upper panel: frame-to-frame $\Delta M / \Delta t$. Lower panel: density RMSE. The solid and dashed curves represent the target and prediction, respectively. The upper time axis displays time in units of the viscous time at the midpoint of the accretion flow, $r=400M$. The lower time axis displays time in units of $10^5 GM/c^3$.} 
\label{fig:direct2}
\end{figure}

In order to assess the accuracy of mass continuity, ideally we would like to compute mass accretion rates. However we are unable to do so because our ML model is unable to predict the velocity field---a necessary ingredient in advection calculations---since it is not present in the training set. The next best thing is to show the time derivative of the mass in the domain. Figure \ref{fig:direct2} shows the frame-to-frame $\Delta M/ \Delta t$ as a function of time. The mass is computed only inside a $r<400 R_S$ box centered on the BH. The predicted mass fluctuations follow closely the target data. The RMSE tells us that the model can predict the following frame with a $\approx 1 \%$ error. This indicates that the DL model nowcasting reproduces quite well the mass variation.

\begin{figure*}
\includegraphics[scale=1.0]{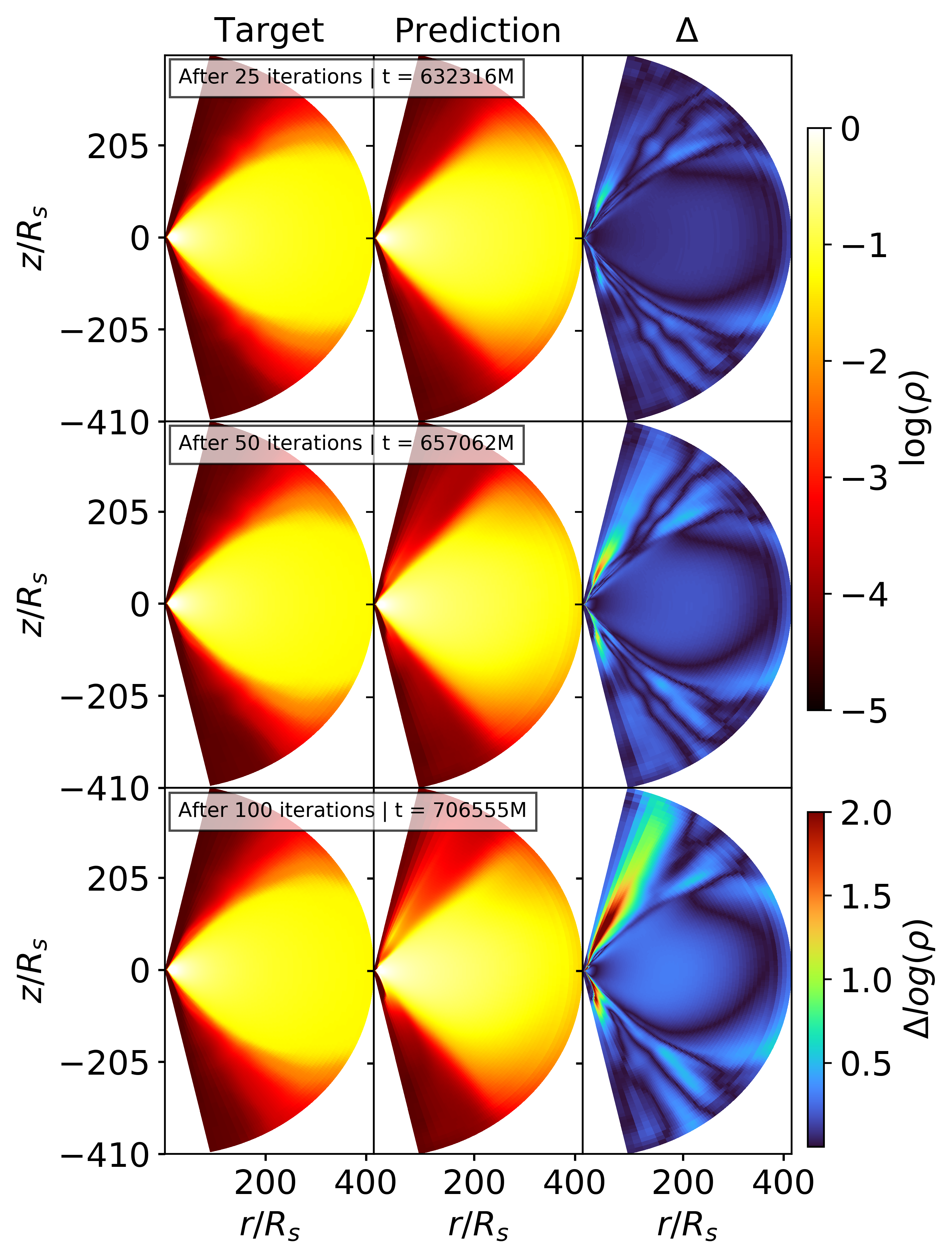}
\caption{Comparison between the target and iterative predictions (one-sim case) at three different times, corresponding to 125, 250 and 500 time steps after the last snapshot of cross-validation at $t =  607570\;M$. } 
\label{fig:iterative1}
\end{figure*}

We assessed the performance of the iterative predictions starting at the first frame of the test set data and then iterating the DL predictions. Figure \ref{fig:iterative1} shows the resulting predictions after $25$, $50$ and $100$ iterations, respectively (or $125$, $250$ and $500$ frames after $t = 607570M$). We see that the general shape and density distribution of the torus itself is preserved in the three cases. However, we see a cumulative discrepancy in the density near the event horizon at $r < 13 R_s$ and along the evacuated funnel with the density increasing in the DL model compared to the target data. This indicates that lower-density regions can be problematic for the DL forecasting. 

\begin{figure}
\centering
\includegraphics[width=\linewidth]{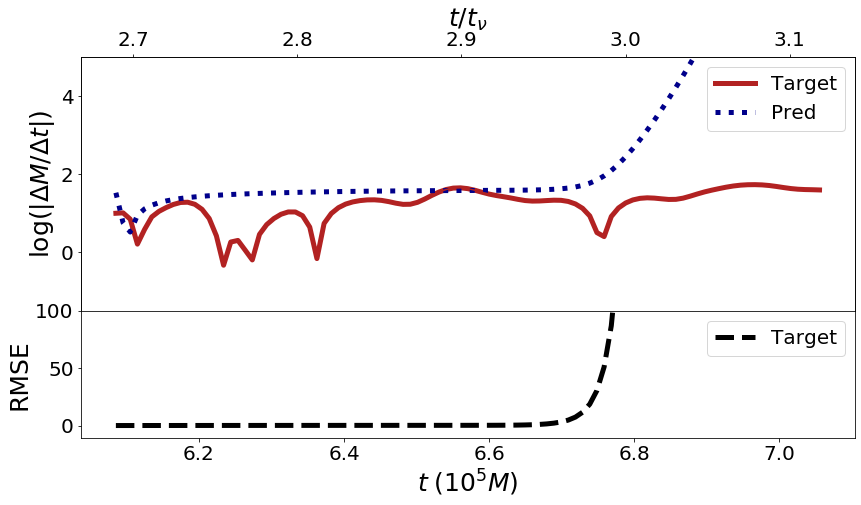}
\caption{Evolution of mass fluctuations based on iterative one-sim model. The same conventions used in Figure \ref{fig:direct2} are adopted here. }
\label{fig:pnss1_deltamdeltat}
\end{figure}

Figure \ref{fig:pnss1_deltamdeltat} shows the mass fluctuations over time for the iterative one-sim case. A couple of points are worth making. First, the model is reproducing the time-averaged mass variation in the domain but fails to capture high-frequency variability (i.e. the valleys in the time series). At $t = 670000M$, the mass starts increasing exponentially as if the model is artificially injecting excess mass in the domain.

Another view of this mass divergence issue is encapsulated in Figure \ref{fig:pnss1_density_mean} which displays the density averaged over the angle $\theta$, as a function of radius and time. Here, we see that the density predicted at small radii diverges around $t = 6.7 \times 10^5 M$. We should note that the torus simulated in PNSS3 does not display considerable variability as can be seen in the constancy over time of the density in the middle panel. In fact, most of the data used in this work for training the neural nets are not dramatically variable. We discuss this limitation in section \ref{sec:disc}.

\begin{figure}
\includegraphics[width=\linewidth]{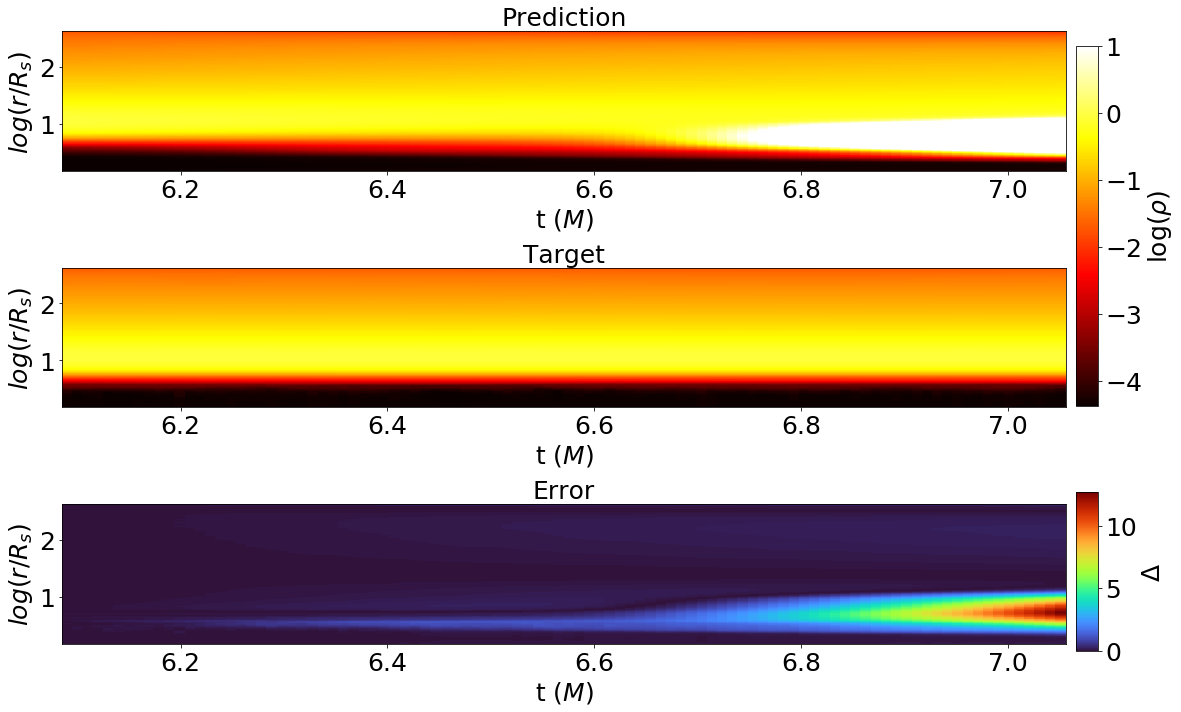}
\caption{Evolution of the density averaged over angle, as a function of radius and time for the iterative one-sim model. The bottom panel shows $\Delta \log \rho$ between target and prediction. } 
\label{fig:pnss1_density_mean}
\end{figure}

\subsection{Model trained on multiple simulations}

Here, we present the results of the multi-sim model which was trained on several hydrodynamical simulations, each with different initial conditions.  

\begin{figure*}
\centering
\includegraphics[width=\linewidth]{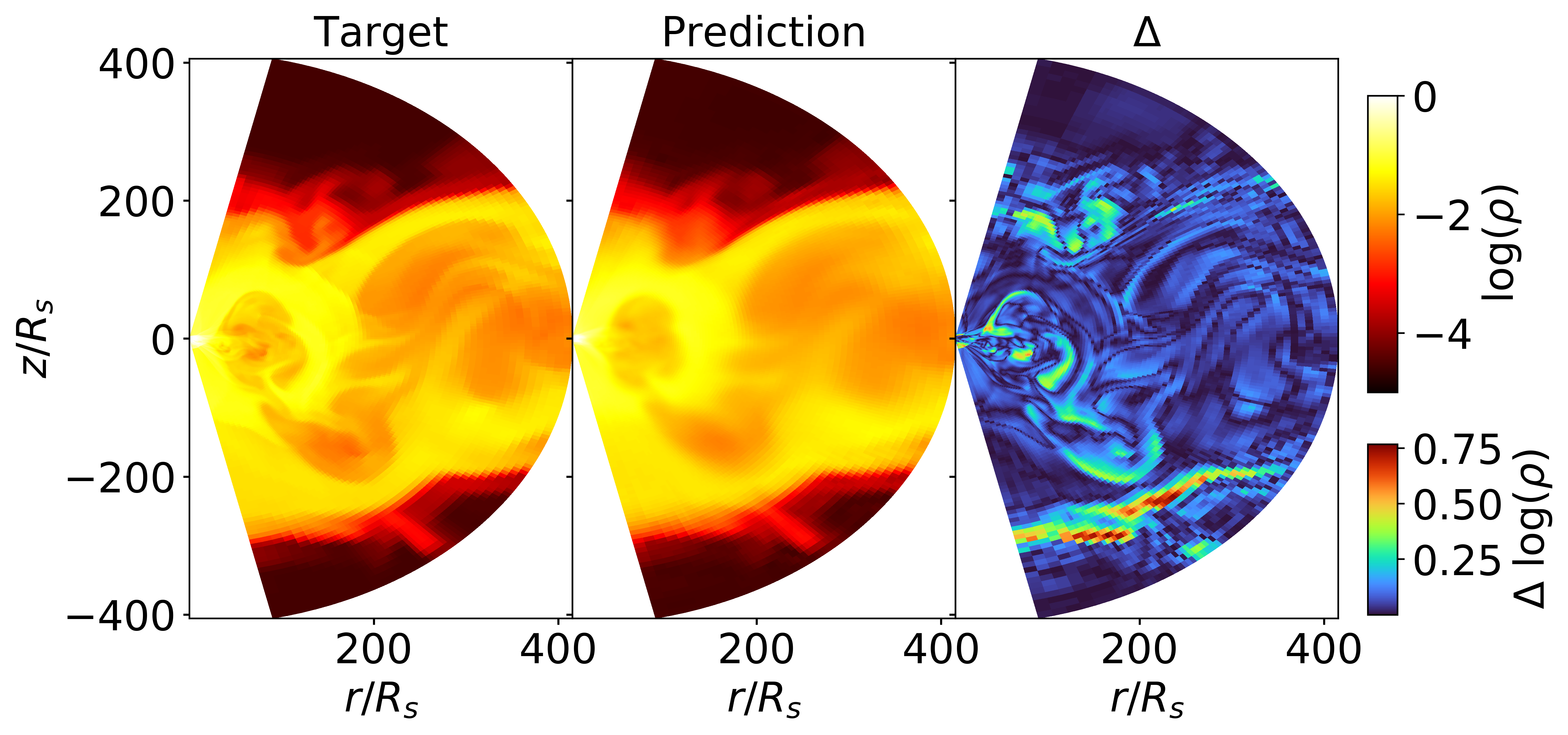}
\caption{Density predictions by the multi-sim direct model, compared with the simulation \code{PNST1} at $t=172035M$.} 
\label{fig:pnst1_direct}
\end{figure*}

We begin by assessing the performance of the multi-sim model's direct prediction for PNST1. Part of the data for this simulation was incorporated in the training of the multi-sim model; Figure \ref{fig:pnst1_direct} shows the model's  prediction for the next frame after being fed as input the density field at $t = 68497M$ from PNST1 which was not used in the training. As can be seen in the figure, the DL model predicts a density field that resembles a spatially smoothed version of the input field; the model successfully reproduces the overall structure of the flow, but fails to capture the small-scale spatial variations of the density. For instance, there is an accumulation of larger residuals (right panel) at the places where there are larger density gradients in the target data.

\begin{figure*}
  \includegraphics[scale=1.0]{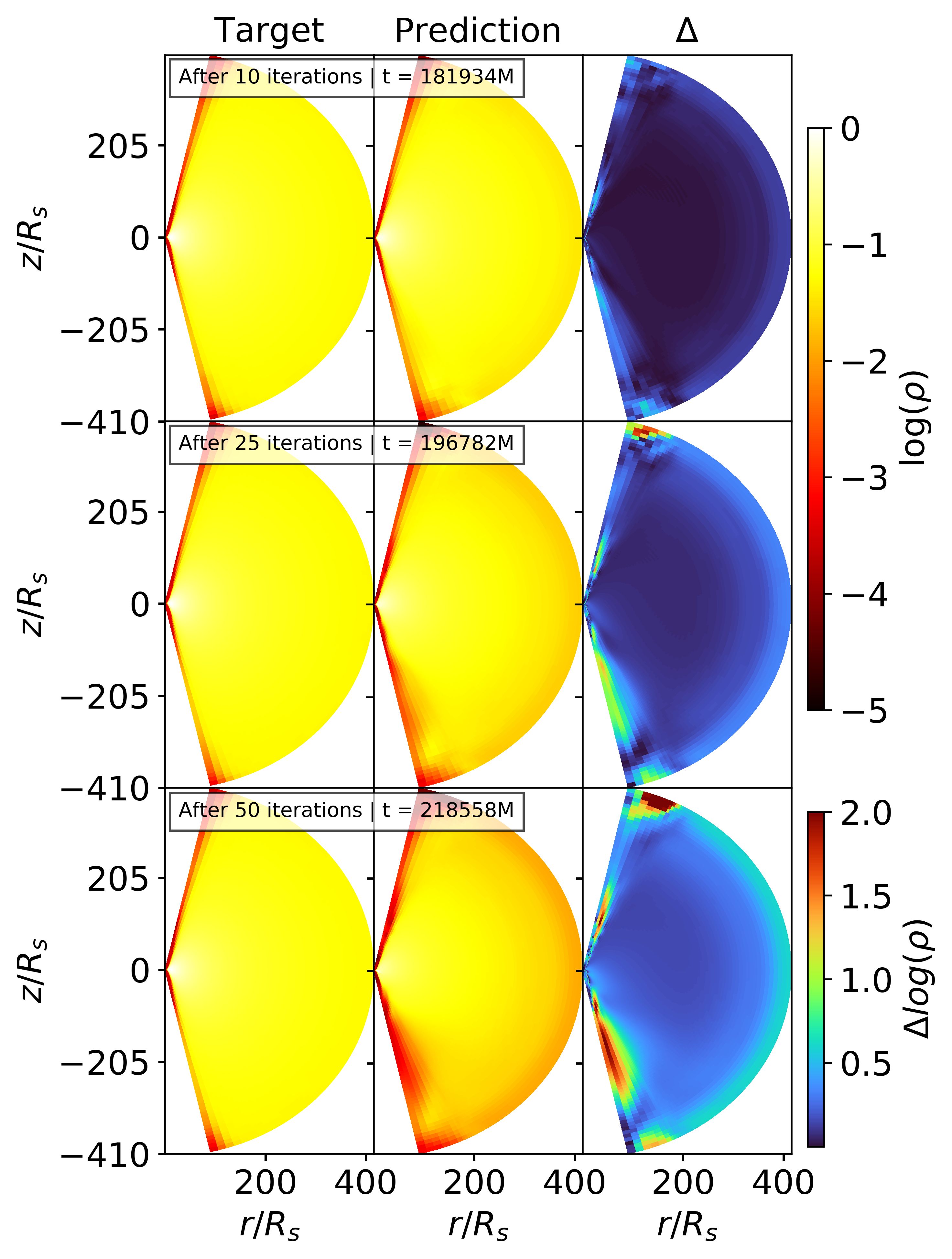}
  \caption{Density predictions by the multi-sim iterative model, compared with the simulation \code{PL0SS3} after 10, 25 and 50 iterations ($t = 181934M$, $196782M$ and $218558M$, respectively).}
  \label{fig:pl0ss3_iterative}
\end{figure*}

Now we use the iterative prediction to assess whether the DL model trained on multiple simulations is able to evolve a an accretion flow consisting of initial conditions that were not present in the training data. We start from the PL0SS3 dataset at $t = 172035M$, waiting until the initial transients of the simulation settle and begin iterating the predictions. Figure \ref{fig:pl0ss3_iterative} shows the results after many iterations, each iteration advancing the system by five frames into the future (the time-difference between two frames is $197.97M$). We see that the predictions increasingly deviate from the target towards the poles as the number of iterations increases, with a density difference between target and prediction of 2 dex, even though the mean difference in the equator remains low ($< 0.1$ dex). We believe this occurs because there is a smaller number of cells near the poles. Due to the lack of training data in those regions, the model does not learn the flow physics well leading to failure modes. 

\begin{figure}
\centering
\includegraphics[width=\linewidth]{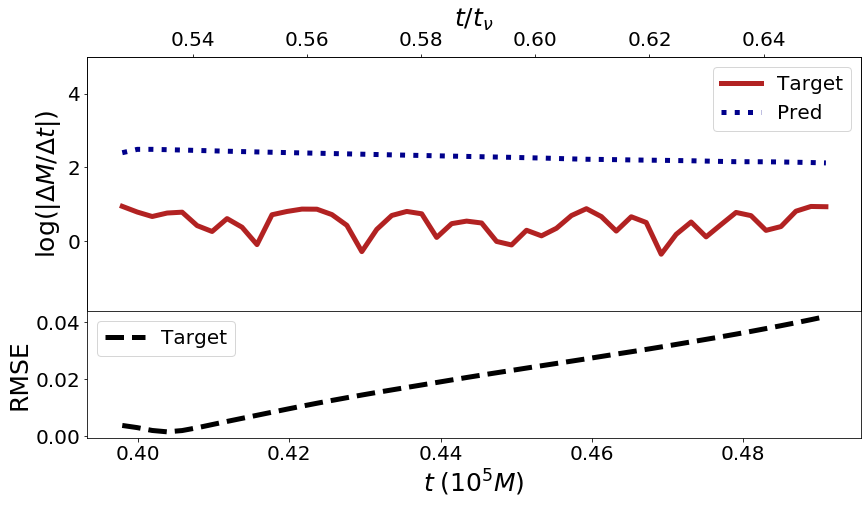}
\caption{Evolution of mass fluctuations based on the iterative multi-sim model for the \code{PL0SS3} dataset. The same conventions used in Figure \ref{fig:direct2} are adopted here.}
\label{fig:pl0ss3_iterative_metrics}
\end{figure}

\begin{figure}
\centering
\includegraphics[width=\linewidth]{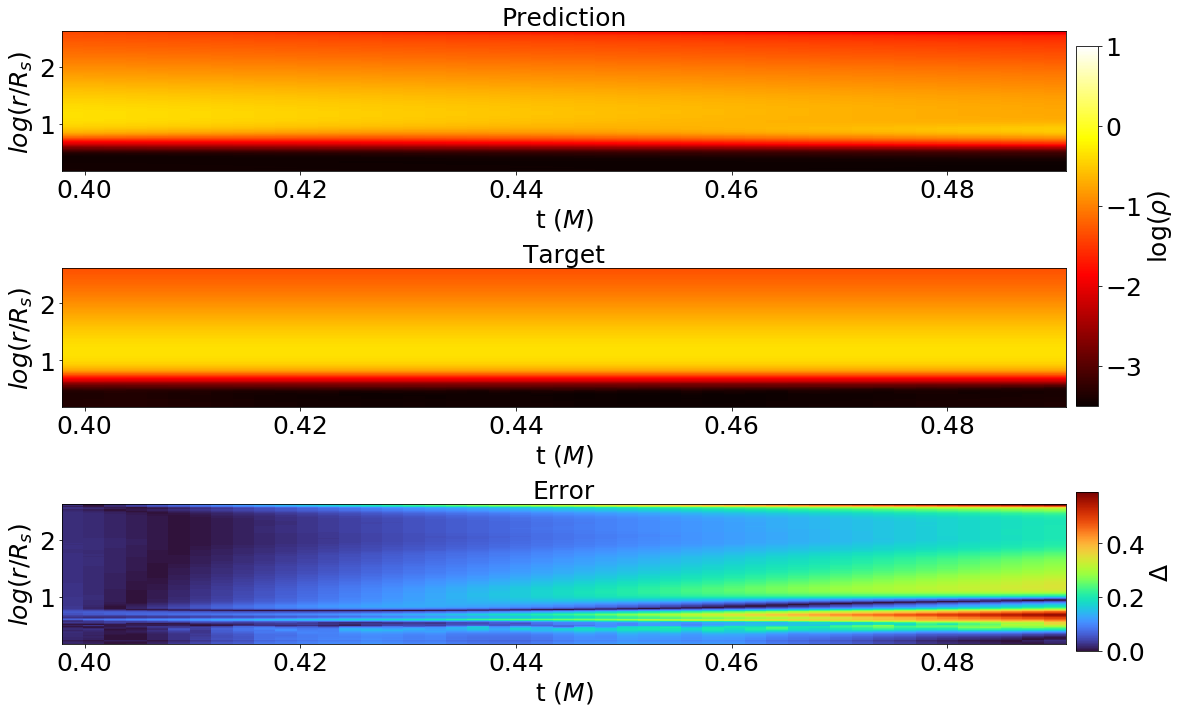}
\caption{Evolution of the density averaged over angle, as a function of radius and time for the iterative multi-sim model, \code{PL0SS3} dataset. The same conventions used in Figure \label{fig:pnss1_density_mean} are adopted here.}
\label{fig:pl0ss3_iterative_mean}
\end{figure}

Figure \ref{fig:pl0ss3_iterative_metrics} displays the mass fluctuations over time for \code{PL0SS3}. Clearly, the predictions shows a large systematic bias towards larger values, by about 1 dex. This indicates that the multi-sim model is having issues with mass continuity, when applied to conditions not present in the training data set. Where is this excess mass injection occurring? Along the poles which are also the region with lower resolution in the training data, as can be seen in Figure \ref{fig:pl0ss3_iterative}. This large density discrepancy is somewhat smoothed out in the $\theta$-average displayed in \ref{fig:pl0ss3_iterative_mean}. 

Figure \ref{fig:pl0ss3_iterative_metrics} also illustrates that the RMSE is increasing linearly with time; the errors build up in the DL model and the predicted density drifts away from the target as seen in Figure \ref{fig:pl0ss3_iterative_mean}, though in the multi-sim model the errors do not diverge exponentially as in the one-sim iterative model. This is good news, since it suggests that training the model with more data reduces the severity of the density divergence.

\section{Discussion}    \label{sec:disc}

Our DL model is based on CNNs and was trained on the solutions of fluid dynamics conservation equations for the problem of an accreting black hole surrounded by a radiatively inefficient accretion flow; the rotating gas is subject to internal viscous stresses, pressure forces and gravity. Our aim was to assess the performance of DL techniques to evolve the spatiotemporal density distribution of a more realistic astrophysical simulation dataset. We considered two different training datasets: the one-sim case where we trained the model on the density spacetime distribution for one single initial condition of the accretion problem, and the multi-sim case where the training set included several different initial conditions. We considered two methods for time-evolving the density distribution provided at a specific time: the direct approach, where the DL model advances only one time-step given a density time-slice (nowcasting), and the iterative approach where the model computes many different future states (forecasting). The iterative approach is particularly relevant since it probes the longer term forecasting capabilities of the trained model.

We begin by discussing the shortcomings of our training data. Most of the simulations used as training data do not display much variability. This is because the models with a larger number of temporal snapshots that we had available to us---which are preferred from the point of view of the data-hungry DL training---are coincidentally those that display less variability. More specifically, in the one-sim case the training data resemble a quasi-laminar, static torus. This is because this specific torus simulation by AN displays weak convective turbulence which translates into little spatial variability of the density\footnote{The exception is the multi-sim model in which about 7 per cent of the $\sim 6000$ temporal snapshots comprising the training data are characterized by larger degrees of variability.}. On one hand these data are interesting because they allow us to evaluate the performance of the DL model under conditions in which the flow is quite laminar, thereby allowing us to separate the variability due to errors introduced by the DL model from intrinsic turbulent variability. The downside is that these data are not representative of realistic accretion flows. Future works should explore training data displaying stronger turbulence and more variability. 

Another shortcoming of the training data is that there is a reduced number of cells as one approaches the poles or towards larger radii. This was implemented by design by AN, because the most important physical phenomena that the authors wanted to study with the simulations did not occur near the poles or at larger radii, so a lower resolution at these regions was acceptable. However, we have found that somehow the subdomains with lower resolution ended up affecting the learning. Concretely, the largest differences between the DL predictions and the target data occurred precisely near the poles or at larger radii. We plan on investigating ways of accounting for nonuniform meshes in the training data in future work.

A third limitation is that our trained DL model is incapable of predicting the turbulent kinetic energy or the turbulence energy spectrum, because the training data only includes the density field as a function of time, not the velocity fluctuations which are required for such estimates. Therefore, we are unable to make any quantitative statements about the turbulence. Including the velocity field in the training set is an interesting future direction.

We have found in the one-sim DL model a strong mass-injection between the polar region and the torus, between $\theta \approx 55^\circ$ and $70^\circ$ (the exact angles depend on the initial conditions of the training dataset). Since our training data do not include velocities, we are unable to say whether this extra mass is outflowing or inflowing. We are also not sure why the neural networks inject vigorous mass in this region. This deserves to be investigated further.

Our long-term goal is quantify the ability of DL models to learn from simulated data of spatiotemporally chaotic astrophysical systems and reproduce the behavior of these systems over time. In this pilot project, we showed that the model could capture the overall spatial features of stable tori which are mostly dynamically stable. In our nowcasting tests the model seem to obey mass continuity reasonably well; in the longer-term forecasting we have had mixed results. The one-sim model can simulate the accretion flow over a duration of $\sim 8 \times 10^4 M$ while reproducing the broad features of mass variability. This duration corresponds to $0.3 t_{\rm visc}$ where $t_{\rm visc}$ is the viscous timescale at $r=400M$, the midpoint of the torus\footnote{$\Delta t = 8 \times 10^4 M$ corresponds to 80 dynamical times at $r=100M$ or about $5 \times 10^3$ dynamical times at $r=6M$}. In the short timescale over which we tested the multi-sim model ($0.1 t_{\rm visc}$), it failed to enforce mass continuity with hot spots near the poles where the DL model injected ten times more mass than in the target data. These are serious issues that deserve further investigation.

\subsection{Speed-up}    \label{sec:speedup}


Here we discuss the potential speedup gained with the DL technique applied to astrophysical fluid dynamic simulations such as those described in this work. Before discussing this topic, however, we should point out two important caveats. 

Before speedup comes \textit{correctness} in numerical calculations. After all, if one accelerates calculations that give incorrect results, one will only arrive faster at the wrong answers. There is  considerable room for improvement in our DL model, since it is not fully capturing mass continuity. Therefore, the DL predictions are not as accurate as the hydrodynamical calculations.

Besides the issue of correctness, one has to be careful not to compare apples and oranges when it comes to the computational time taken by different approaches. Whereas the training data was generated by a code that solves the relevant fluid conservation equations and computes the velocity, pressure and density fields as a function of time, the DL model outputs a time-varying image of the density field; no dynamical information is provided because it was not incorporated in the training. Whereas the hydrodynamic code solves hyperbolic partial differential equations, the DL model computes gradients (back-propagation), convolutions and locates the minimum of a multidimensional surface. Therefore, we have two sets of completely different numerical methods. This is an important reason for taking the comparisons of wall times taken by such methods with a grain of salt.

Keeping these caveats firmly in mind, it can be nevertheless instructive to quantify the speedup achieved by the DL method. If for nothing else, it gives us a rough idea of the potential acceleration that can be achieved by ML techniques applied to BH accretion flows, even though their accuracy still needs to be significantly improved. 

Our first comparison takes into account that the hydrodynamic simulations are evolving in time more information than the DL learning method. The hydrodynamical model returns four fields (two components of velocity, pressure and density) on a $400 \times 200$ mesh every 168 seconds for the PNST1 case on a CPU cluster with 200 cores, whereas the DL model computes the density field for the flow configuration on a $256 \times 192$ mesh every 0.01 s on a GPU\footnote{Given the way the CNN is structured, it actually outputs density fields at five sequential moments, all at once. However, here we consider this as a single physical field output.}. Thus, the hydro and DL models output $\approx 2$ and $\sim 2 \times 10^4$ floats per millisecond of physical information, respectively. The DL is outputting physical information $\sim 10^5$ faster than the hydro calculations. For other data sets, the speedups are equally dramatic as shown in Table \ref{TableSpeedUp}.

\begin{table*}
\begin{center}
\begin{tabular}{|c|c|c|c|c|c|c|}
\hline \hline
 Name & $\Delta t\;(M)$ & Wall time (s) & Wall time (s) & Floats/ms & Floats/ms & Speed Up \\ [0.5ex] 
  &  & Prediction & Target & Prediction & Target & \\
 \hline\hline
 PNSS3 & 64341 & 4 & 54600 & 3994 & 0.381 & 10483  \\
 PNST1 & 198 & 0.01 & 168 & 24576 & 1.905 & 12902 \\
 PL0SS3 & 9305 & 0.47 & 7896 & 25099 & 1.945 & 12902  \\
 \hline
\end{tabular}
\caption{Comparison of the performance to simulate different tori setups using hydrodynamical simulations versus DL. \textbf{Column contents:} 1. Name of the setup. 2. duration of simulation in units of $M$. 3. Wall time in seconds taken by the hydrodynamical simulation to generate 4x400x200 meshes (four fluid properties: $P$, $\rho$, $v_1$, $v_2$). 4. Wall time in seconds taken by the DL model to generate 1x256x192 mesh (one fluid property: $rho$). 5. Floats of fluid properties generated per millisecond by the hydrodynamical simulation. 6. Floats of fluid properties generated per millisecond by the DL model. 7. Speedup achieve to compute fluid floats by the DL model compared to the fluid simulation (i.e. ratio of column 5 to column 6. }
\label{TableSpeedUp}
\end{center}
\end{table*}

\begin{figure}
\includegraphics[width=\linewidth]{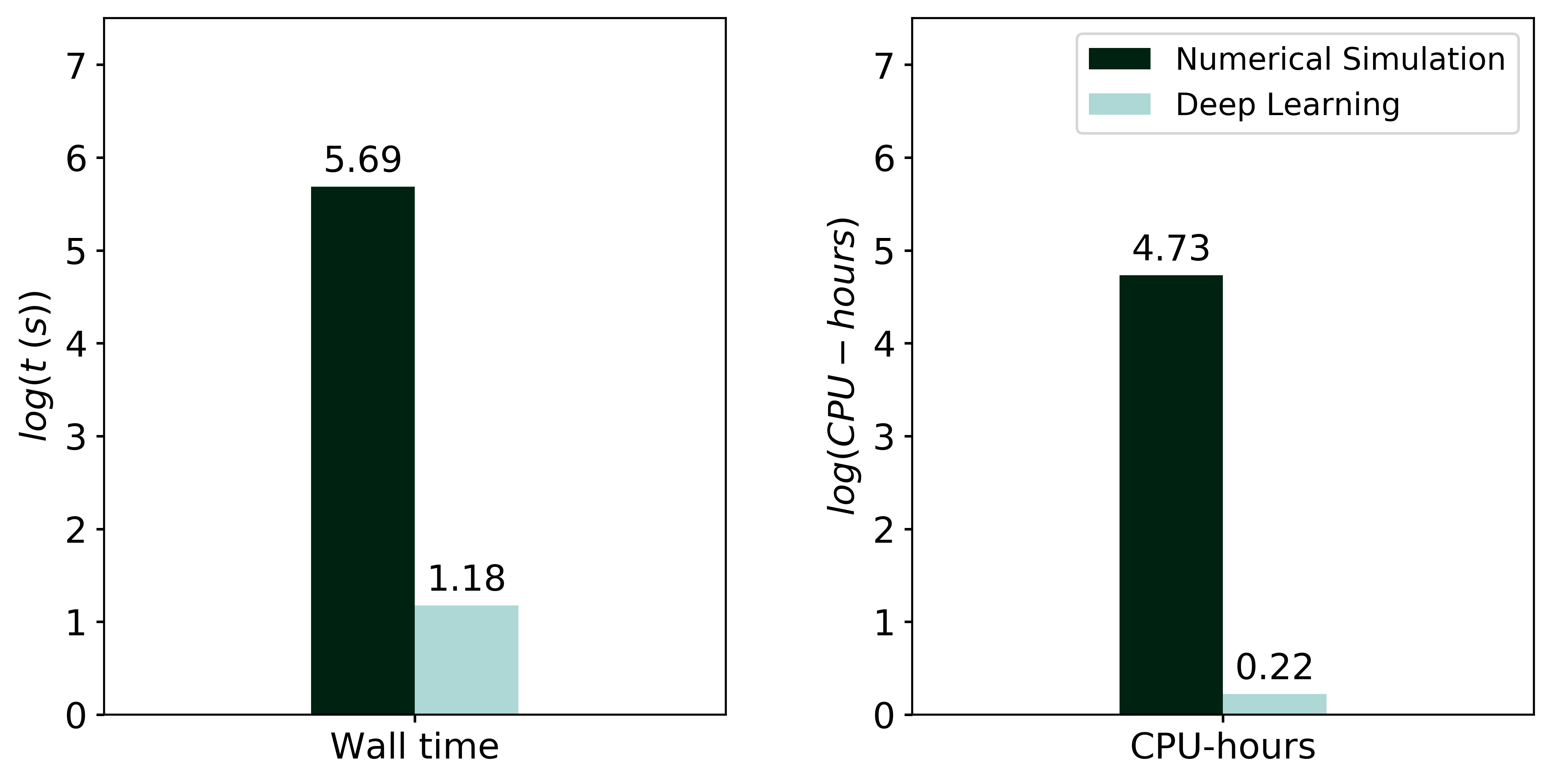}
\caption{Comparison of the wall time to evolve the accretion flow using the iterative method (one-sim) in seconds (left panel) and CPU-hours (right panel). Shorter is better in this plot. The black rectangle indicates the performance of traditional numerical simulations based on the Godunov approach implemented on a CPU cluster, whereas the light green rectangle shows the performance of the DL model inference on a single GPU. The DL model was previously trained on two GPUs.}
\label{fig:speedup_1}
\end{figure}

Figure \ref{fig:speedup_1} compares the wall time and CPU-hours taken by the hydrodynamical simulation and the DL model to advance the state of the accreting BH from $t=6.1 \times 10^5 M$ to $6.8 \times 10^5 M$ (one-sim, iterative case). As is tradition in the ML literature, these estimates do not take into account the time required to generate the training data sets \eg{Pfaff2020, Li2021}. We have an overall speedup of a factor of $3.2 \times 10^4$ times. However, when taking into account the training time, the speedup is decreased to a factor of $38$. This is the ratio between the wall time taken by the hydro code to evolve the torus (PNSS3 dataset) over a duration of $\Delta t=7\times 10^4 M$ and the corresponding wall time taken by the trained DL model, including the time required for training and not considering the time required to generate the training set. 

\begin{figure}
\includegraphics[width=\linewidth]{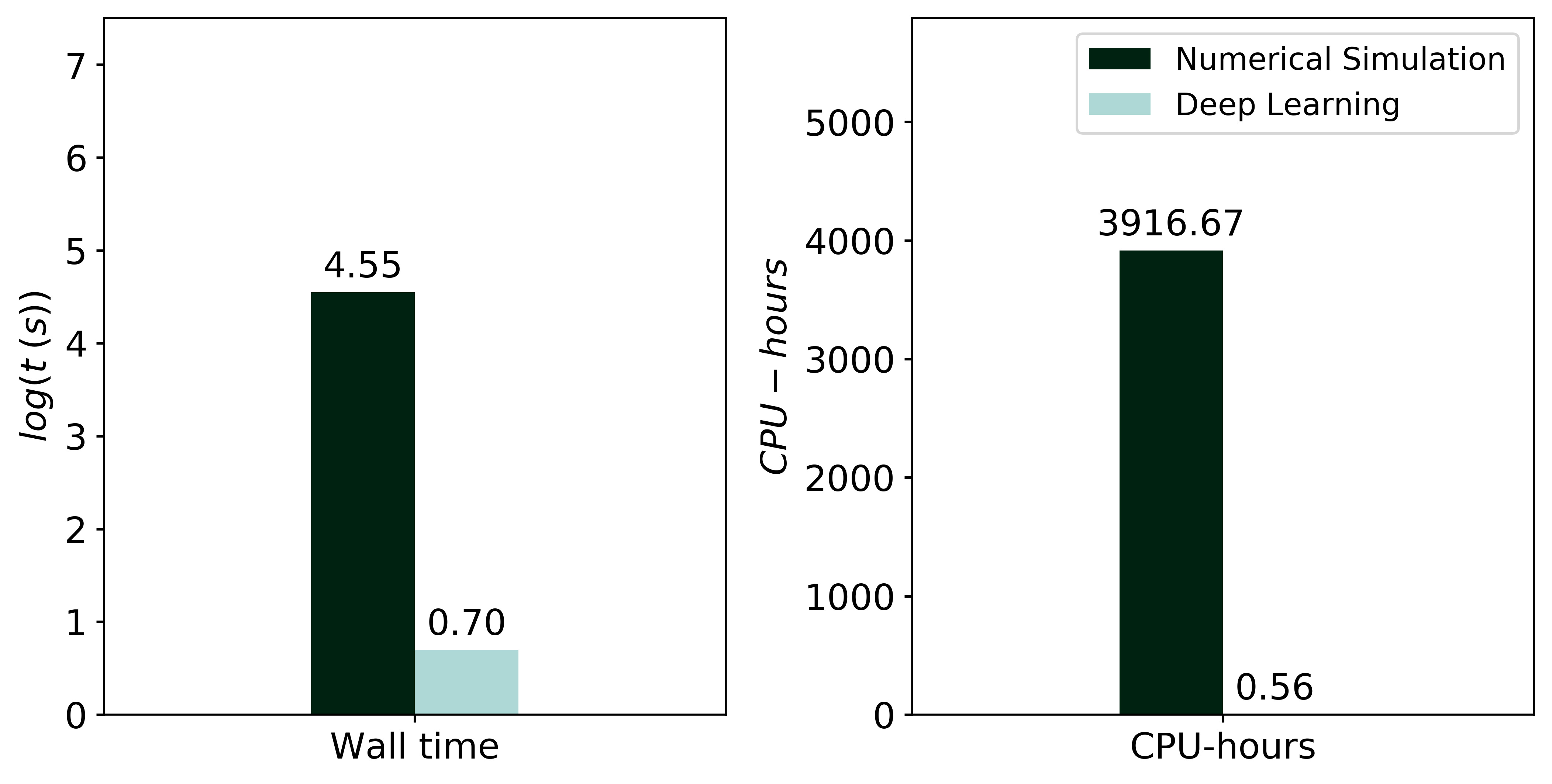}
\caption{Performance charts for the multi-sim iterative deep learning model (PL0SS3 dataset) versus standard fluid dynamics solvers. We follow the same conventions as in Figure \ref{fig:speedup_1}. }
\label{fig:speedup_2}
\end{figure}

Figure \ref{fig:speedup_2} compares the time taken to evolve the PL0SS3 simulation using standard fluid dynamics numerical methods versus DL techniques using the multi-sim, iterative approach. We obtain a factor of about $7 \times 10^3$ speed-up with the DL model with respect to conventional CPU fluid dynamics solvers. 

\section{Conclusions}   \label{sec:conclusion}

In this pilot study we have trained a machine to make black hole weather forecasting, by using a deep learning model to evolve in time a spatiotemporally chaotic astrophysical system consisting of a accreting black hole which feeds on a large gas reservoir. The training dataset consists of the numerical solutions to the hydrodynamical Newtonian equations for a range of initial conditions. The setup corresponds to a Schwarzschild black hole surrounded by a radiatively inefficient accretion flow which extends from 2 to 400 Schwarzschild radii. Our main conclusions can be summarized as follows: 

(i) We find that convolutional neural networks trained on a single simulation of a torus in quasi-equilibrium predict reasonably well the overall spatiotemporal density distribution of the flow over a duration of $8\times 10^4 GM/c^3$, which corresponds to $0.3$ viscous times at $400 M$.

(ii) The above DL model reproduces well the time-averaged mass variation in the computational domain but fails to capture high-frequency variability.

(iii) The reality imagined by the deep learning model drifts from the training dataset over time---an ``artificial Alzheimer''. In this case, mass is artificially injected and the error increases to $\sim 90$ per cent after $8\times 10^4 GM/c^3$. 

(iv) When we train the DL model with several simulations of accretion flows spanning multiple initial conditions, the resulting model has only moderate success evolving an accretion flow with initial conditions not present in the training dataset for a duration of $8000\; GM/c^3$. Even though the DL model reproduces in a broad-brush sense the density spatial distribution, it violates mass continuity. 

(v) The DL model seems to ``learn too well'' from the training dataset. This results in artifacts in the regions of the flow near the poles where the mesh contains less cells, since our hydrodynamical simulations are based on a nonuniform grid. At such regions, we observe an injection of mass by the DL model which can be attenuated by training the CNN with more models. 
    
(vi) Keeping in mind that our DL model can only predict density distributions at the moment, once trained it evolves on a single GPU an accretion flow $10^4$ times faster than traditional numerical fluid dynamics integrators running on 200 CPUs. 

The caveats of the work are mostly related to limitations in the training data, as follows: (a) most of the data consist of a torus with little variability, (b) the data corresponds to purely hydrodynamic simulations whereas black hole accretion is an inherently magnetized phenomenon, (c) the data has a limited resolution and is purely two-dimensional and (d) we only considered density fluctuations in the training. Future investigations should improve on these aspects.

In conclusion, our results indicate that deep learning models are a promising way for evolving black hole accretion flows but they still have a long way to go. If in the future DL models achieve a forecasting accuracy comparable to traditional fluid solvers while maintaining the speed gains reported here, they could bring about a revolution in numerical studies of accretion physics.

We believe that a data-driven machine learning approach holds promise for accelerating not only fluid dynamics simulations, but also general relativistic magnetohydrodynamic ones. The recent development of physics-informed deep learning and physics-inspired neural networks is worth mentioning. For instance, one approach resembles traditional fluid dynamical solvers \citep{ruiwang2020}. \cite{ruiwang2020} introduced convolutional neural networks to replace spatial filtering and temporal average while solving turbulent flows. The application of physics-inspired neural networks can follow conservation laws given by the fluid equations. A potential future application may combine lagrangian neural networks with convolutional neural networks as a predictor. Lagrangian neural networks can parametrize lagrangians from observed or simulated data using neural networks \citep{cranmer2020}; they were designed to respect conservation laws. An alternative approach is using hybrid models to create robust predictors such as the U-Net combined with discriminators \citep{eskimez2021} and a variational autoencoder combined with a generative model \citep{cheng2020}. Adding a discriminator can make the model more robust since it learns the loss function from the training data. Other promising lines of work include mesh-grid simulations using graph neural networks \citep{Pfaff2020} and vision transformers \citep{girdhar2021}. These topics certainly deserves further investigations.

 \section*{Acknowledgements}

We acknowledge productive discussions with Ivan Almeida, Fabio Cafardo, Gustavo Soares, Nando de Freitas, Mike Walmsley, Leandro Kerber, Amelie Saintonge, Françoise Combes and Andrew Humphrey. We also acknowledge the useful discussions that occurred during the Khipu 2019 and UK-Brazil Frontiers of Science 2020 workshops. We thank the anonymous referee for the feedback that increased the scientific quality of this work. R. D. was supported by CAPES (Coordenação de Aperfeiçoamento de Pessoal de Nível Superior) Proex. R. N. was supported by FAPESP (Funda\c{c}\~ao de Amparo \`a Pesquisa do Estado de S\~ao Paulo) under grant 2017/01461-2. The Black Hole Group received the donation of two GPUs from NVIDIA: a Quadro P6000 under the GPU Grant Program and a GP100 from NVIDIA Brasil.

\bibliographystyle{mnras}
\bibliography{refs,refs-Nemmen,AI} 

\appendix

\section{Architecture}
\label{sec:arch}

\begin{figure}
\centering
\includegraphics[scale=0.35]{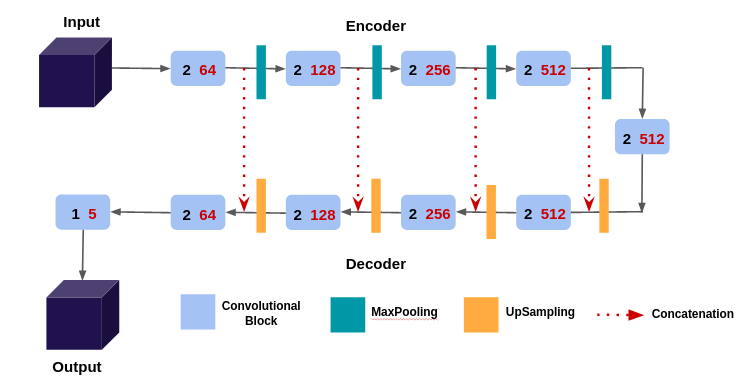}
\caption{Illustration of the UNet architecture used to train the model.}
\label{fig:unet}
\end{figure}

The neural network architecture UNet \citep{Ronneberger2015} that is used in this paper is shown in Table \ref{TableappB} and illustrated in Figure \ref{fig:unet}. 

The encoder is composed of four blocks, and each block has two convolutional layers and a MaxPooling layer. The encoder receives the input and transforms it, generating intermediate feature maps with lower dimensionality in $(x,y)$ plane. The latent space, which is also convolutions, connects the encoder to the decoder. The decoding operations transform the generated feature maps, gradually upsampling it until the output tensors reach the same dimension of inputs. There are skip connections between the encoder outputs and the decoder inputs. The concatenations connect the information of the encoder with the decoder. In this way, the decoder will base their decision on the primary information of the encoder. We feed our network with the training set and use the validation set to Early-Stopping procedure \citep{Prechelt1996}.

\begin{table}
\begin{center}
\begin{tabular}{|c|c|c|}
\hline
 Name & Operation & Output Shape \\ [0.5ex] 
 \hline\hline
 Input & $\;$ & $(N, 256, 192, 5)$ \\ 
 
 Conv-1-1 & Conv2D + ReLU & $(N, 256, 192, 64)$ \\ 
 
 Conv-1-2 & Conv2D + ReLU & $(N, 256, 192, 64)$  \\
 
 Max-1 & MaxPooling2D & $(N, 128, 96, 64)$  \\
 
 Conv-2-1 & Conv2D + ReLU & $(N, 128, 96, 128)$ \\ 

 Conv-2-2 & Conv2D + ReLU & $(N, 128, 96, 128)$  \\
 
 Max-2 & MaxPooling2D & $(N, 64, 48, 128)$  \\

 Conv-3-1 & Conv2D + ReLU & $(N, 64, 48, 256)$ \\ 
 
 Conv-3-2 & Conv2D + ReLU & $(N, 64, 48, 256)$  \\

 Max-3 & MaxPooling2D & $(N, 32, 24, 256)$  \\
 
 Conv-4-1 & Conv2D + ReLU & $(N, 32, 24, 512)$ \\ 
 
 Conv-4-2 & Conv2D + ReLU & $(N, 32, 24, 512)$  \\
 
 Max-4 & MaxPooling2D & $(N, 16, 12, 256)$  \\

 Conv-5-1 & Conv2D + ReLU & $(N, 16, 12, 512)$ \\ 
 
 Conv-5-2 & Conv2D + ReLU & $(N, 16, 12, 512)$  \\
 
 UpSam-1 & UpSampling2D & $(N, 32, 24, 512)$ \\ 
 
 Conc-1 & Conc(UpSam-1, Conv-4-2) & $(N, 32, 24, 1024)$  \\
 
 Conv-6-1 & Conv2D + ReLU & $(N, 32, 24, 512)$ \\ 
 
 Conv-6-2 & Conv2D + ReLU & $(N, 32, 24, 512)$  \\
 
 UpSam-2 & UpSampling2D & $(N, 64, 48, 512)$ \\ 
 
 Conc-2 & Conc(UpSam-2, Conv-3-2) & $(N, 64, 48, 768)$  \\
 
 Conv-7-1 & Conv2D + ReLU & $(N, 64, 48, 256)$ \\ 
 
 Conv-7-2 & Conv2D + ReLU & $(N, 64, 48, 256)$   \\
 
 UpSam-3 & UpSampling2D & $(N, 128, 96, 256)$ \\ 
 
 Conc-3 & Conc(UpSam-3, Conv-2-2) & $(N, 128, 96, 384)$  \\
 
 Conv-8-1 & Conv2D + ReLU & $(N, 128, 96, 128)$ \\ 
 
 Conv-8-2 & Conv2D + ReLU & $(N, 128, 96, 128)$  \\
 
 UpSam-4 & UpSampling2D & $(N, 256, 192, 128)$ \\ 
 
 Conc-3 & Conc(UpSam-4, Conv-1-2) & $(N, 256, 192, 192)$  \\
 
 Conv-9-1 & Conv2D + ReLU & $(128, 256, 192, 64)$ \\ 
 
 Conv-9-2 & Conv2D + ReLU & $(128, 256, 192, 64)$  \\
 
 Output & Conv2D + ReLU & $(N, 256, 192, 5)$  \\
 \hline
\end{tabular}
\caption{{\label{TableappB}}
Details of the architecture UNet we used in this project. Conc is concatenation and $N$ is the batch size.}
\end{center}
\end{table}

We performed all the DL experiments using two NVIDIA Quadro GPUs from Pascal architecture, GP100 and P6000. The implementation was done in \code{Keras} \citep{chollet2015keras} v2.1  with the \code{TensorFlow} \citep{tensorflow2015} v1.8 backend. 

\section{Data preparation}
\label{sec:dataprep}

To avoid a biased model, we performed data preparation. The data preparation consists of normalize the density values, crop the grid of the simulations, and create 4D--arrays that will serve as input and output of our network. First, we normalize the density values, in the range $[0,1]$ using a logarithm normalization:

\begin{equation}
\label{eq:eq4}
       \rho_{NORM} =  \frac{log(\rho) - log(min(\rho))}{log(max(\rho)) - log(min(\rho))}.
\end{equation}

\noindent The normalization avoids bias by putting all values in the same scale. In our raw data, the range goes from  $10^{-5}$ up to  $10^1$, without normalization the largest values might dominated. Normalization helps as well to speed-up the learning, converging faster \citep{Sola1997}.  

The next step of the data preparation is the crop of our grid. The crop was performed as follows: in the radial direction, we removed 144 cells that span $\,9500\,GM/c^2$, encompassing mostly atmosphere with $\rho < 10^{-3}$. Meanwhile, in the polar direction, we remove eight cells that correspond to $\sim 3.5^\circ$ along the poles. We performed the crop after initial tests where the atmosphere dominated during the learning procedure. Our main interest is to evaluate how much the model can learn the accretion flow dynamics so removing the atmosphere does not present major drawbacks. 

\begin{figure}
\centering
\includegraphics[scale=0.40]{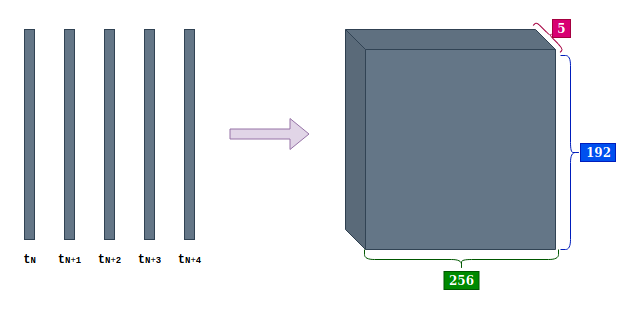}
\caption{We append five consecutive ($256 \times 192$) arrays creating a $(N \times 256 \times 192 \times 5)$ array with $N$ being the number of the snapshots. $t_N$ is the density field associated with the $n_{th}$ snpashot.}
\label{fig:tensor}
\end{figure}

The final part is to build our blocks that will feed the network. Since we want to forecast density fields after a $\Delta t$, we build the blocks to incorporate temporal information. The scheme in Fig. \ref{fig:tensor} shows how we build the blocks. We attach five consecutive density field creating a block -- $N \times 256 \times 192 \times 5$ -- with $N$ being the number of data. $N$ is defined after the point of the accretion rate becomes stationary. 

\section{Fluid dynamic simulations details}
\label{sec:simconfig}

We summarize the properties --- kinematic viscosity $\nu$, Shakura-Sunyaev's parameter $\alpha$, and the angular momentum $\ell (r)$ --- of nine simulations in Table \ref{TableappA}. AN explore two parametrizations of the kinematic viscosity. 



To our interests, we will distinguish the angular momentum profile as PN or PL and the viscosity profile $\nu$ as ST or SS. The time difference between two snapshots is $\Delta t = 197.97M$. 

\begin{table}
\begin{center}
\begin{tabular}{|c|c|c|c|c|}
\hline
 Name & $\ell(R)$ & $\nu$ & $\alpha $ & $\times 10^5\;T\;(M)$ \\ [0.5ex] 
 \hline\hline
 PNST01 & PN & ST & 0.01 & 8.0  \\
 
 PNST1 & PN & ST & 0.1 & 0.9 \\
 
 PNSS1 & PN & SS & 0.1 & 4.5 \\
 
 PNSS3 & PN & SS & 0.3 & 3.3 \\
 
 PL0ST1 & PL & ST & 0.1 & 0.8 \\
 
 PL0SS3 & PL & SS & 0.3 & 2.1 \\
 
 PL2SS1 & PL & SS & 0.1 & 1.4 \\
 
 PL2SS3 & PL & SS & 0.3 & 2.1 \\
 \hline
\end{tabular}
\caption{{\label{TableappA}}
The initial configurations of each simulation: angular momentum profile, viscosity profile, and the alpha parameter. In the last column, we show the duration of each simulation.}
\end{center}
\end{table}

\section{``All systems" case}
\label{allsimproc}

\begin{figure}
\includegraphics[scale=0.50]{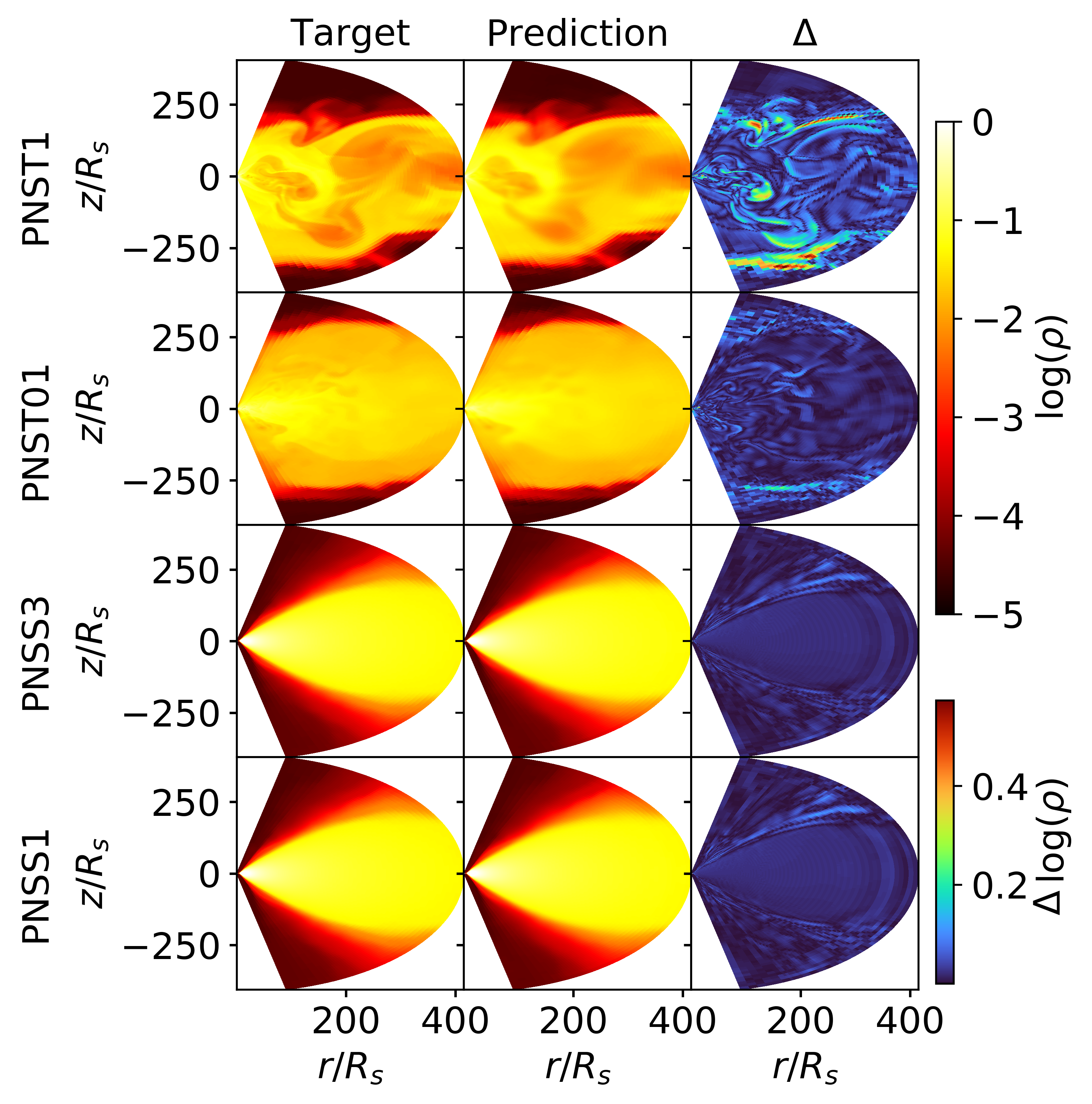}
\caption{We show the 10th step after the training set ends for all simulations. Each simulation has different gravitational time in the 10th step, but the $\Delta t = 1979.7M $ is the same for all of them. The density profile of the target and the prediction as well as the difference $\Delta$ plot of four systems. The PNSS3 system is the one in the ``one simulation" case.}
\label{fig:appendix_pred1}
\end{figure}

\begin{figure}
\includegraphics[scale=0.55]{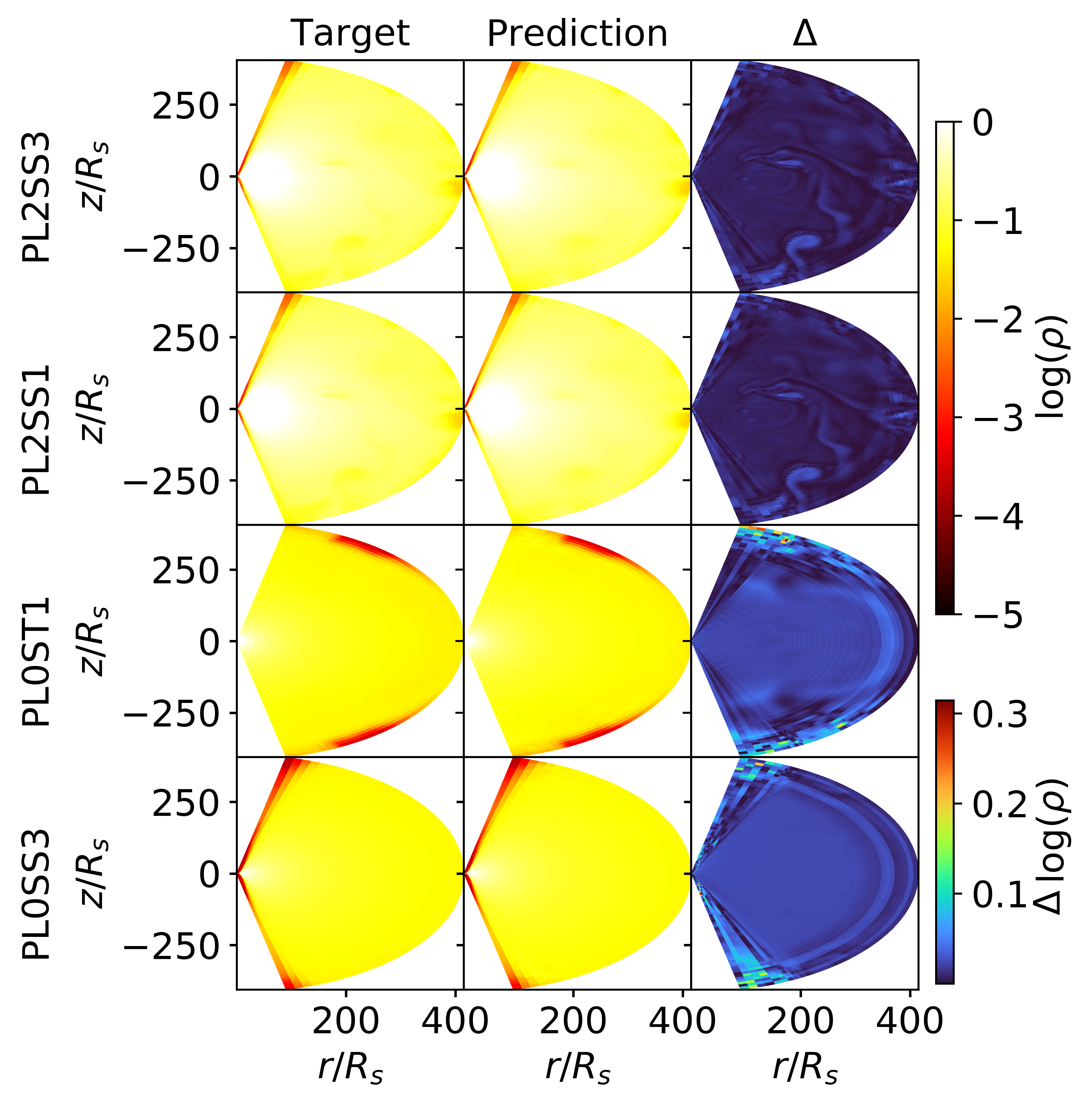}
\caption{The same as in figure \ref{fig:appendix_pred1} for the rest of simulations. We show further analysis to the PL0SS3 simulation since the model was not trained with this configuration.}
\label{fig:appendix_pred2}
\end{figure}

Figure \ref{fig:appendix_pred1} and figure \ref{fig:appendix_pred2} shows the direct predictions of all simulations equivalent to $\Delta t = 1979.7M$ in the future. We see that the model manages to predict all simulations, as in ``one simulation" case, if we feed the previous snapshot of the simulation. 

\bsp	
\label{lastpage}
\end{document}